\documentclass[lettersize,journal]{IEEEtran}
\usepackage[IEEEtran]{./research17}

\usepackage{cite}
\usepackage{amsmath,amssymb,amsfonts}
\usepackage{algorithmic}
\usepackage{graphicx}
\usepackage{textcomp}
\usepackage{xcolor}
\def\BibTeX{{\rm B\kern-.05em{\sc i\kern-.025em b}\kern-.08em
    T\kern-.1667em\lower.7ex\hbox{E}\kern-.125emX}}
    
\usepackage[numbers,sort&compress]{natbib}
\usepackage{algorithm}
\usepackage{algorithmic}

\usepackage{fancyhdr}

\usepackage{graphicx}
\usepackage{float}
\usepackage{subfig}

\makeatletter
\def\thanks#1{\protected@xdef\@thanks{\@thanks
		\protect\footnotetext{#1}}}
\makeatother

\begin{document}
\title{Secrecy Capacity Analysis and Beamforming Optimization for MIMO-VLC Wiretap Channels}

\author{Sufang~Yang,
	Longguang~Li,~\IEEEmembership{Member,~IEEE,}
	Jintao~Wang,~\IEEEmembership{Senior~Member,~IEEE,}
	Ya Li, Liang Xia, Hongjun He, Qixing Wang, and Guangyi Liu,~\IEEEmembership{Member,~IEEE}
	\thanks{{This work was supported in part by the National Natural Science Foundation of China under Grant No. 62101192, in part by Shanghai Sailing Program under Grant No. 21YF1411000, and in part by Tsinghua University-China Mobile Research Institute Joint Innovation Center}. \emph{(Corresponding author: Longguang Li.)}}
	\thanks{{Sufang~Yang, Ya Li, Liang Xia, Hongjun He, Qixing Wang, and Guangyi Liu are with the Future Research Laboratory, China Mobile Research Institute, Beijing 100053, China (e-mail: yangsufang@chinamobile.com).}}
	\thanks{Longguang Li is with SJTU Paris Elite Institute of Technology, Shanghai Jiao Tong University, Shanghai 200240, China (e-mail: llg9012@sjtu.edu.cn).}
	\thanks{Jintao~Wang is with the Beijing National Research Center for Information Science and Technology (BNRist), Tsinghua University, Beijing 100084, China, also with the Department of Electronic Engineering, Tsinghua University, Beijing 100084, China, and also with the Research Institute, Tsinghua University in Shenzhen, Shenzhen 518057, China (e-mail: wangjintao@tsinghua.edu.cn).}
}

\date{}

\maketitle
\begin{abstract}
This paper investigates a multiple-input multiple-output (MIMO) visible light communication (VLC) wiretap channel consisting of a transmitter, a legitimate receiver, and an eavesdropper. The optical input is subject to both peak- and average-intensity constraints. By applying the generalized entropy-power inequality to truncated exponential inputs, we derive a novel closed-form expression for the achievable secrecy rate for general MIMO VLC configurations. To enhance transmission confidentiality, a fully-connected beamforming scheme is proposed, along with a low-complexity sub-connected alternative. Although the resulting beamforming design problems are non-convex, they are efficiently addressed by transforming them into a sequence of convex subproblems solvable via the successive convex approximation framework. Numerical results demonstrate that the proposed schemes achieve significant secrecy performance improvements compared with the benchmark scheme.

\end{abstract}


\section{Introduction}
With the accelerated evolution of intelligent transportation systems, vehicular networks place increasingly stringent requirements on wireless communications for vehicle-to-vehicle, vehicle-to-infrastructure, and vehicle-to-everything links. In particular, emerging vehicular applications necessitate the simultaneous provision of high throughput, ultra-low latency, massive connectivity, and high reliability. Such requirements expose the inherent limitations of conventional radio-frequency (RF) communication systems. Specifically, the scarcity of available spectrum has led to severe congestion in RF-based vehicular networks. To address this challenge, alternative wireless communication technologies are urgently needed to complement existing RF frameworks. Visible light communication (VLC) has emerged as a promising solution by exploiting the abundant spectrum in the 400-800 THz range \cite{Zeng2009,Kaushal2017,Chi2020}. In VLC systems, information is modulated onto visible light and transmitted using light-emitting diodes (LEDs), serving dual functions for illumination and communication. The modulated light propagates through the optical wireless channel and is detected by photodetectors (PDs) at the receiver.

{Owing to the broadcast nature of wireless transmission, any receiving node within the coverage area is capable of intercepting the transmitted signal, which creates a significant security risk when potential eavesdroppers (illegal receivers) coexist with legitimate receivers -- a scenario modeled as a wiretap channel. To mitigate such security risks, physical-layer security (PLS) has been proposed to enhance confidentiality for legitimate transmissions by degrading the eavesdropper's channel. Although extensive studies have established PLS results for RF wiretap channels ~\cite{Spencer2004,Zhang2012,Zou2013,Ng2014,Liu2017}, they cannot be directly extended to the VLC wiretap channels due to their fundamentally different channel characteristics. The key difference stems from the fact that VLC adopts the intensity-modulation direct-detection scheme. Thus, the input of the LED in VLC channels is real and nonnegative, in sharp contrast to the complex-valued input in RF channels. Moreover, the input is subject to a peak-intensity constraint to ensure that the LED operates within its linear range \cite{Peng2021,Amini2023,Lapidoth2009}, as well as an average intensity constraint imposed by illumination requirements or safety considerations \cite{Hammadi2021,Wang2013}. Motivated by these distinctive characteristics, a growing body of research has investigated the secrecy capacity of VLC wiretap channels under PLS, which is defined as the maximum achievable secrecy rate at which the transmitter can securely communicate with the legitimate receiver in the presence of an eavesdropper.


For the single-input single-output (SISO) VLC wiretap channels, existing studies have characterized the secrecy capacity under various optical input constraints \cite{Mostafa2015,Zaid2015,Arfaoui2016-globecom,Arfaoui2018-1,Wang2018,Al-Sallami2025}. Under a single peak-intensity constraint, achievable secrecy rate is derived by employing uniformly distributed input, which naturally yields a lower bound on the secrecy capacity \cite{Mostafa2015}. This lower bound is shown to be tight at high signal-to-noise ratio (SNR) but loose at low SNR. To further tighten the secrecy capacity bounds, alternative input distributions have been investigated, including truncated Gaussian distribution \cite{Zaid2015}, truncated generalized normal distribution \cite{Arfaoui2016-globecom}, and truncated discrete generalized normal distribution \cite{Arfaoui2018-1}. Beyond the single peak-intensity constraint, the secrecy capacity analysis has also been extended to scenarios with an average-intensity constraint alone, as well as combined peak- and average-intensity constraints, for which corresponding upper and lower bounds on the secrecy capacity are derived \cite{Wang2018,Al-Sallami2025}.


For the multiple-input single-output (MISO) wiretap channels, existing research focuses on enhancing the achievable secrecy rate through many PLS techniques. Representative approaches include beamforming \cite{Mostafa2015,Mostafa2016}, friendly jamming (or artificial noise) \cite{Mostafa2014-ICC,Mostafa2014-globecom}, and space modulation (SM) \cite{Wang2018-Tcom,Wang2019}. Under the peak-intensity constraint and perfect eavesdropper channel state information (CSI), a beamforming scheme is proposed to maximize the achievable secrecy rate \cite{Mostafa2015}. To tackle the issue arising from imperfect eavesdropper CSI, a robust beamforming scheme is developed to maximize the worst-case achievable secrecy rate within the eavesdropping region \cite{Mostafa2016}. Alongside beamforming, friendly jamming effectively degrades the eavesdropper performance. Some studies assume that each LED simultaneously transmits an information-bearing signal and a randomly generated jamming signal \cite{Mostafa2014-ICC}, whereas others introduce a separate jammer dedicated solely to transmitting the jamming signal \cite{Mostafa2014-globecom}. Furthermore, an SM scheme is leveraged to improve performance for the legitimate receiver. Different LED activation strategies are investigated, where a subset of LEDs is activated at any given time to transmit the signal \cite{Wang2018-Tcom}, whereas only a single LED is activated at a time \cite{Wang2019}. Building upon these individual techniques, cooperative PLS schemes that jointly exploit the aforementioned techniques have also been explored to further enhance secrecy performance \cite{Arfaoui2020-surv,Su2021,Hsiao2019,Yang2022,Wu2025}.

For the multiple-input multiple-output (MIMO) VLC wiretap channels, several studies have extended the MISO results to this scenario \cite{Arfaoui2017-PIMRC,Minh2014,Chen2017,Arfaoui2020-twc}. Nevertheless, these approaches still exhibit notable limitations. Specifically, the works in \cite{Minh2014} and \cite{Chen2017} neglect essential optical input constraints. In contrast, \cite{Arfaoui2017-PIMRC} considers both peak- and average-intensity constraints; however, it relies on inappropriate modeling assumptions by treating the average intensity (i.e., the first-order moment) as a second-order moment, which contradicts fundamental characteristics of VLC signals \cite{Lapidoth2009,Wang2013,Hammadi2021}. Furthermore, although the work in \cite{Arfaoui2020-twc} correctly incorporates the peak-intensity constraint to derive the achievable secrecy rate\footnote{{In \cite{Arfaoui2020-twc}, the authors consider that the input $\mathbf{X}$ is subject to both peak- and average-intensity constraints such that $\mathrm{Pr}\{|X_i|\leq \amp\}=1$ and $\textsf{E}\{X_i\}=0$, $\forall i\in\{1,\cdots,n_\textnormal{T}\}$. However, it can be shown that in \eqref{eq: alpha_i}, $\alpha_i=\frac{1}{2}$ for all $i$, indicating that the average-intensity constraint is inactive \cite[Lem. 2]{Lapidoth2009}, \cite[Lem. 1]{Chaaban2016} \cite[Prop. 1]{Moser2017}. Consequently, the analysis in \cite{Arfaoui2020-twc} effectively corresponds to the scenario in which a single peak-intensity constraint is imposed on the input.}}, its formulation is computationally intensive, involves multiple undetermined design parameters, and is restricted to this single type of input constraint.

{To the best of our knowledge, the MIMO-VLC wiretap channels under both peak-intensity and average-intensity constraints have not yet been systematically investigated. Consequently, a concise and tractable closed-form expression for the achievable secrecy rate is still lacking, which is of fundamental importance for both theoretical analysis and practical system design. The main contributions are summarized as follows.
\begin{itemize}
	\item By applying the generalized entropy-power inequality (GEPI) to the truncated exponential input, we derive closed-form achievable secrecy rates for the MIMO-VLC wiretap channels. Different quantitative relationships between the number
	of transmit and receive apertures are discussed; see Theorem~\ref{thm1}.
	\item Two novel beamforming schemes, termed fully-connected and sub-connected beamforming schemes, are proposed to enhance the PLS of the MIMO-VLC wiretap channel. The optimal beamformer for each scheme is formulated by an optimization problem maximizing the derived achievable secrecy rate. The conventional zero-forcing (ZF) beamforming scheme is also analyzed as a benchmark; see Sec.~\ref{sec: beamforming design}.
	\item To reduce the complexity of finding the optimal beamformer, we use the Taylor approximation to transform the original non-convex beamforming optimization problem into a sequence of convex subproblems. Novel algorithms are then proposed to iteratively solve these subproblems and find the optimal beamformer that maximizes the achievable secrecy rate; see Algorithms~\ref{alg: alg1}, \ref{alg: alg2}, and \ref{alg: alg3}.
\end{itemize}

The remaining part of this paper is organized as follows. Sec. ~\ref{sec: channel model} introduces the MIMO-VLC wiretap channel model. Sec.~\ref{sec: secrecy rate} formulates the achievable secrecy rates. Secs.~\ref{sec: beamforming design} and~\ref{sec: algorithm} propose beamforming schemes and the corresponding algorithms. Simulation results are provided in Sec.~\ref{sec: simulations}. We conclude this paper in Sec.~\ref{sec: conclusion}. A few proofs are given in the Appendix.

\emph{Notation:} Scalars are denoted by non-bolded letters, e.g., $X$ denotes a random scalar and $x$ its realization. Vectors are denoted by boldfaced letters, e.g., $\mathbf{X}$ denotes a random vector and $\mathbf{x}$ its realization. All the matrices in this paper are deterministic and typeset in a blackboard-bold font, e.g., $\mathbb{H}$. Mutual information is denoted by $\textsf{I}(\cdot;\cdot)$, entropy by $\textsf{h}(\cdot)$, expectation by $\textsf{E}\{\cdot\}$, and covariance by $\textsf{Cov}\{\cdot\}$. The absolute determinant of matrices is denoted by $|\cdot|$, the transposition by $(\cdot)^\mathsf{T}$, the trace by $\mathrm{tr}\{ \cdot \}$, and the vectorization by $\mathrm{vec}(\cdot)$. $1$-norm of matrices is denoted by $\|\cdot\|_1$, and the Frobenius norm by $\|\cdot\|_\mathsf{F}$. The set of real numbers is denoted by $\mathcal{R}$, nonnegative real numbers by $\mathcal{R}_+$, and $n$-dimensional positive semi-definite matrices by $\mathcal{S}_{+}^{n}$. $\mathbb{I}_{m}$ denotes an $(m\times m)$-dimensional identity matrix, and $\mathbf{e}_j$ denotes a unit vector with the $j$-th entry equal to $1$. $\mathbf{0}_m$ and $\mathbf{1}_m$ denote column vectors of length $m$ with all entries equal to $0$ and $1$, respectively. $\mathbf{0}_{m\times n}$ and $\mathbf{1}_{m\times n}$ denote $(m\times n)$-dimensional matrices with all entries equal to $0$ and $1$, respectively.  



\section{Channel Model}\label{sec: channel model}

Consider an MIMO-VLC wiretap channel consisting of a transmitter Alice, a legitimate receiver Bob, and an eavesdropper Eve. Alice aims to communicate messages to Bob while preventing eavesdropping by Eve. We use subscripts $(\cdot)_\textnormal{B}$ and $(\cdot)_\textnormal{E}$ to distinguish parameters associated with Bob and Eve, respectively. Assume Alice has $n_\textnormal{T}$ LEDs, while Bob and Eve have $n_\textnormal{B}$ and $n_\textnormal{E}$ PDs, respectively. A fixed direct current bias, $\mathrm{I}_\textnormal{DC}\in\mathcal{R}_+$, is introduced to ensure that the LED's input is nonnegative \cite{Ma2017}. As it can be effectively removed by a capacitor at the receiver, the channel outputs observed at Bob and Eve are characterized by
\begin{IEEEeqnarray}{rCl}
	\subnumberinglabel{channel model}
	\mathbf{Y}_\textnormal{B} &=& \xi \mathbb{H}_\textnormal{B} \mathbf{X} + \mathbf{Z}_\textnormal{B},\\
	\mathbf{Y}_\textnormal{E} &=& \xi \mathbb{H}_\textnormal{E} \mathbf{X} + \mathbf{Z}_\textnormal{E},	
\end{IEEEeqnarray}
with the photoelectric coefficient $\xi\in\mathcal{R}_+$, the channel matrices $\mathbb{H}_\textnormal{B}\in\mathcal{R}^{n_\textnormal{B} \times n_\textnormal{T}}_+$ and $\mathbb{H}_\textnormal{E}\in\mathcal{R}^{n_\textnormal{E} \times n_\textnormal{T}}_+$, the data input $\mathbf{X}\in\mathcal{R}^{n_\textnormal{T}}$, the outputs $\mathbf{Y}_\textnormal{B}\in\mathcal{R}^{n_\textnormal{B}}$ and $\mathbf{Y}_\textnormal{E}\in\mathcal{R}^{n_\textnormal{E}}$, and the noises $\mathbf{Z}_\textnormal{B}\in\mathcal{R}^{n_\textnormal{B}}$ and $\mathbf{Z}_\textnormal{E}\in\mathcal{R}^{n_\textnormal{E}}$, following Gaussian distributions $\mathcal{N}( \mathbf{0}_{n_\textnormal{B}},  \sigma_\textnormal{B}^2 \mathbb{I}_{ n_\textnormal{B} })$ and $\mathcal{N}( \mathbf{0}_{ n_\textnormal{E} },  \sigma_\textnormal{E}^2 \mathbb{I}_{ n_\textnormal{E} } )$, respectively. Without loss of generality, we assume $\xi=1$ and $\sigma_\textnormal{B}^2=\sigma_\textnormal{E}^2=1$. Throughout the paper, we assume that Alice knows Eve's CSI \cite{Arfaoui2020-surv}. We also assume that $\mathbb{H}_\textnormal{B}$ and $\mathbb{H}_\textnormal{E}$ are of full ranks, consistent with the setup in existing works \cite{Moser2017,Li2020,Chaaban2018} \footnote{As demonstrated in \cite[Sec. III-C]{Moser2017} and \cite[Appendix A]{Li2020}, any channel model can be reduced via singular value decomposition (SVD) to an equivalent channel model with a full-rank channel matrix​. This simplification streamlines the secrecy capacity analysis without loss of generality.}. 
%

Considering the limited dynamic range and illumination requirement of LEDs \cite{Peng2021,Amini2023,Lapidoth2009,Hammadi2021,Wang2013}, the input $\mathbf{X}$ is subject to the following constraints:
\begin{itemize}
	\item {Peak-intensity constraint}: 
	\begin{IEEEeqnarray}{rCl}
		\mathrm{Pr} \{ |X_i| \leq \amp \} = 1, \ \forall i \in\{1,\cdots,n_\textnormal{T}\}, \label{eq: peak cons}
	\end{IEEEeqnarray}
	\item {Average-intensity constraint}:
	\begin{IEEEeqnarray}{rCl}
	\textsf{E}\{X_i\} = \EE_i, \ \forall i \in\{1,\cdots,n_\textnormal{T}\}, \label{eq: ave cons}
	\end{IEEEeqnarray}
\end{itemize}
where $X_i$ denotes the $i$-th entry of $\mathbf{X}$. For convenience, we define the vector $\bm{\alpha} =(\alpha_1,\cdots,\alpha_{n_\textnormal{T}})^\mathsf{T}$ with
\begin{IEEEeqnarray}{rCl}
	\alpha_i=\frac{\EE_i+\amp}{2\amp}, \ \forall i \in\{1,\cdots,n_\textnormal{T}\}. \label{eq: alpha_i}
\end{IEEEeqnarray}
It is direct to see $\alpha_i\in(0,1)$ \cite{Lapidoth2009}. We also define the vector $\bm{\beta}=(\beta_1,\cdots,\beta_{n_\textnormal{T}})^\mathsf{T}$ with $\bm{\beta} =\bm{\alpha} - \frac{1}{2} \mathbf{1}_{n_\textnormal{T}}$.

In this wiretap channel, the secrecy capacity $\mathsf{C}_\textnormal{s}$, denoting the maximum achievable secrecy rate between Alice and Bob in the presence of Eve, is characterized by the following lemma:
\begin{lemma}[{\cite[Cor. 2]{Csiszár1978}}, {\cite[Thm. 22.1]{Gamal2011}}]\label{lem1}
	The secrecy capacity for the wiretap channel in \eqref{channel model} is given by	
	\begin{IEEEeqnarray}{rCl}
		\mathsf{C}_\textnormal{s} = \max_{ p_{ \mathbf{U},\mathbf{X} } } \bigl\{ \textsf{I}( \mathbf{U};\mathbf{Y}_\textnormal{B} ) - \textsf{I}( \mathbf{U};\mathbf{Y}_\textnormal{E} ) \bigr\}, \label{eq: lemma}
	\end{IEEEeqnarray}
	where $\mathbf{U}$ is an auxiliary variable and $\mathbf{U} \rightarrow \mathbf{X} \rightarrow ( \mathbf{Y}_\textnormal{B},\mathbf{Y}_\textnormal{E} ) $ form a Markov chain.
	$\hfill\blacksquare$
\end{lemma}

\section{Achievable Secrecy Rate Characterization}\label{sec: secrecy rate}
This section presents the achievable secrecy rate of the MIMO-VLC wiretap channel. By setting $\mathbf{U}=\mathbf{X}$, the secrecy capacity can be lower-bounded as
\begin{IEEEeqnarray}{rCl}
	\mathsf{C}_\textnormal{s} &\geq&  \max_{ p_{ \mathbf{X} } } \{ \textsf{I}( \mathbf{X};\mathbf{Y}_\textnormal{B} ) - \textsf{I}( \mathbf{X};\mathbf{Y}_\textnormal{E} ) \}. \label{eq: Cs bound1}
\end{IEEEeqnarray}
{To satisfy the constraints in \eqref{eq: peak cons} and \eqref{eq: ave cons}, we employ a truncated exponential distribution $\mathbf{X}_\textnormal{te}$, which performs well in a scalar point-to-point VLC channel and achieves channel capacity at high SNR \cite{Lapidoth2009,Moser2017,Li2020}}. Its distribution is given by
\begin{IEEEeqnarray}{rCl}
	p_{\mathbf{X}_\textnormal{te}}(\mathbf{x}) = \prod_{i=1}^{n_\textnormal{T}} \frac{1}{2\amp} \cdot
	\frac{\mu_i}{1-e^{-\mu_i}} e^{-\frac{\mu_i(x_i+\amp)}{2\amp}}, \ x_i\in[-\amp,\amp], \quad \label{eq: texp pdf}
\end{IEEEeqnarray}
where $x_i$ denotes the $i$-th entry of $\mathbf{x}$, and $\mu_i$ is the unique solution to 
\begin{IEEEeqnarray}{rCl}
	\frac{1}{\mu_i} - \frac{e^{-\mu_i}}{1-e^{-\mu_i}} = \alpha_i, \ i\in\{1,\cdots,n_\textnormal{T}\}.
\end{IEEEeqnarray}
Substituting \eqref{eq: texp pdf} into \eqref{eq: Cs bound1}, the secrecy capacity in \eqref{eq: Cs bound1} can be further lower-bounded as
\begin{IEEEeqnarray}{rCl}
	\mathsf{C}_\textnormal{s} &\geq&  \textsf{I}( \mathbf{X}_\textnormal{te};\mathbf{Y}_\textnormal{B} ) - \textsf{I}( \mathbf{X}_\textnormal{te};\mathbf{Y}_\textnormal{E} ). \label{eq: Cs bound2}
\end{IEEEeqnarray}
It is worth noting that \eqref{eq: Cs bound2} can be efficiently analyzed based on quantitative relationships between the number of {transmit and receive apertures} \cite{Moser2017,Li2020,Chaaban2018}. We classify the secrecy capacity into two primary cases in this paper\footnote{{The analysis for (a) $n_\textnormal{T} \geq n_\textnormal{B}$ and $n_\textnormal{T} < n_\textnormal{E}$, (b) $n_\textnormal{T} < n_\textnormal{B}$ and $n_\textnormal{T} \geq n_\textnormal{E}$, follow similar procedures as in Secs. \ref{sec: Cs lb case i} and \ref{sec: Cs lb case ii}, which are omitted here.}}:
\begin{itemize}
	\item Case I: the number of LEDs is larger than or equal to the number of PDs, i.e., $n_\textnormal{T} \geq n_\textnormal{B}$ and $n_\textnormal{T} \geq n_\textnormal{E}$;
	\item Case II: the number of LEDs is less than the number of PDs, i.e., $n_\textnormal{T} < n_\textnormal{B}$ and $n_\textnormal{T} < n_\textnormal{E}$.
\end{itemize}
We summarize the main results in the following theorem.
\begin{theorem}\label{thm1}
	For Case I, an achievable secrecy rate is given by:
	\begin{IEEEeqnarray}{rCl}
	\mathsf{R}_\textnormal{s}^\textnormal{I}(\mathbb{H}_\textnormal{B},\mathbb{H}_\textnormal{E})	 
	&=& \frac{n_\textnormal{B}}{2} \log \left( 1 + \left| \mathbb{H}_\textnormal{B} \cdot \diag{ \mathbf{p} } \cdot \mathbb{H}_\textnormal{B}^\mathsf{T} \right|^{\frac{1}{n_\textnormal{B}}} \right) \nonumber\\
	&& -\frac{1}{2} \log \left|\mathbb{H}_\textnormal{E} \cdot \diag{ \mathbf{v} } \cdot \mathbb{H}_\textnormal{E}^\mathsf{T}  + \mathbb{I}_{n_\textnormal{E}}\right|, \label{eq: Cs lb case i} \qquad\quad\
	\end{IEEEeqnarray}
	for Case II:
	\begin{IEEEeqnarray}{rCl}
	\mathsf{R}_\textnormal{s}^\textnormal{II}(\mathbb{H}_\textnormal{B},\mathbb{H}_\textnormal{E})
	&=& \frac{n_\textnormal{T}}{2} \log \left( 1 + \left| \diag{\mathbf{p}} \right|^{\frac{1}{ n_\textnormal{T} } } \cdot \left|  \mathbb{H}_\textnormal{B}^\mathsf{T} \mathbb{H}_\textnormal{B}  \right|^{\frac{1}{n_\textnormal{T}}} \right) \nonumber\\
	&& -\frac{1}{2} \log \left| \diag{\mathbf{v}} \cdot \mathbb{H}_\textnormal{E}^\mathsf{T} \mathbb{H}_\textnormal{E} +  \mathbb{I}_{ n_\textnormal{T} } \right|, \label{eq: Cs lb case ii}  \qquad\qquad
	\end{IEEEeqnarray}
	where vectors $\mathbf{p}=(p_1,\cdots,p_{n_\textnormal{T}})^\mathsf{T}$ and $\mathbf{v}=(v_1,\cdots,v_{n_\textnormal{T}})^\mathsf{T}$, with $\forall i\in\{1,\cdots,n_\textnormal{T}\}$,
	\begin{IEEEeqnarray}{rCl}
		p_i &=& 2\amp^2 \frac{  e^{2\alpha_i \mu_i}  }{\pi e} \left( \frac{1-e^{-\mu_i}}{\mu_i}\right)^2 ,\label{eq: p_vec}\\
		v_i &=& \amp^2 \left( \frac{8}{\mu_i^2} - \frac{4e^{-\mu_i}+4}{\mu_i(1-e^{-\mu_i})} -4\alpha_i^2 + 4 \alpha_i \right). \label{eq: g2 function} \quad
	\end{IEEEeqnarray}	
\end{theorem}

The proofs of \eqref{eq: Cs lb case i} and \eqref{eq: Cs lb case ii} are given in Secs.~\ref{sec: Cs lb case i} and \ref{sec: Cs lb case ii}, respectively.
$\hfill\blacksquare$


\subsection{Proof of Achievable Secrecy Rate for Case I}\label{sec: Cs lb case i}
We first derive a lower bound on $\textsf{I}( \mathbf{X}_\textnormal{te};\mathbf{Y}_\textnormal{B} )$ and then an upper bound on $\textsf{I}( \mathbf{X}_\textnormal{te};\mathbf{Y}_\textnormal{E} )$. The lower bound derivation begins with
\begin{IEEEeqnarray}{rCl}
	\textsf{I}( \mathbf{X}_\textnormal{te};\mathbf{Y}_\textnormal{B} ) 
	&=& \textsf{h}( \mathbb{H}_\textnormal{B}  \mathbf{X}_\textnormal{te} + \mathbf{Z}_\textnormal{B} ) - \textsf{h}( \mathbf{Z}_\textnormal{B} ) \\
	&\geq& \frac{n_\textnormal{B}}{2} \log \left( 1 + \frac{ e^{ \frac{2}{ n_\textnormal{B} } \textsf{h}( \mathbb{H}_\textnormal{B}  \mathbf{X}_\textnormal{te} ) } }{ e^{ \frac{2}{ n_\textnormal{B} } \textsf{h}( \mathbf{Z}_\textnormal{B} ) } } \right) \label{eq: 38}\\
	&\geq& \frac{n_\textnormal{B}}{2} \log \left( 1 + \left| \mathbb{H}_\textnormal{B} \cdot \diag{ \mathbf{p} } \cdot \mathbb{H}_\textnormal{B}^\mathsf{T} \right|^{\frac{1}{n_\textnormal{B}}} \right), \label{eq: I(s_te;y_b) lower bound1}
\end{IEEEeqnarray}
where \eqref{eq: 38} holds by the entropy power inequality (EPI) and \eqref{eq: I(s_te;y_b) lower bound1} by the GEPI \cite{Zamir1993}:
\begin{IEEEeqnarray}{rCl}
	\textsf{h}( \mathbb{H}_\textnormal{B}  \mathbf{X}_\textnormal{te} )
	&\geq& \frac{n_\textnormal{B}}{2} \log \left( 2\pi e \left| \mathbb{H}_\textnormal{B} \cdot \diag{ \mathbf{p} } \cdot \mathbb{H}_\textnormal{B}^\mathsf{T} \right|^{\frac{1}{n_\textnormal{B}}} \right), \label{eq: 19} \qquad
\end{IEEEeqnarray}
where $\mathbf{p} = \left( \frac{e^{2\textsf{h}( X_{\textnormal{te},1} ) } }{2\pi e} , \cdots, \frac{e^{2\textsf{h}( X_{ \textnormal{te},n_\textnormal{T} } ) }}{2\pi e} \right)^\mathsf{T}$ with $X_{\textnormal{te},i}$ denoting the $i$-th entry of $\mathbf{X}_{\textnormal{te}}$, $i\in\{ 1,\cdots,n_\textnormal{T} \}$. Then, we compute the entropy $\textsf{h}( X_{ \textnormal{te},i} )$ using \eqref{eq: texp pdf}, and obtain the vector $\mathbf{p}$ in \eqref{eq: p_vec}.

Next, an upper bound on $\textsf{I}(\mathbf{X}_\textnormal{te};\mathbf{Y}_\textnormal{E} )$ can be obtained by
\begin{IEEEeqnarray}{rCl}
	\textsf{I}( \mathbf{X}_\textnormal{te};\mathbf{Y}_\textnormal{E} ) 	
	&=& \textsf{h}( \mathbb{H}_\textnormal{E}  \mathbf{X}_\textnormal{te} + \mathbf{Z}_\textnormal{E} ) - \textsf{h}( \mathbf{Z}_\textnormal{E} )\\
	&\leq& \frac{n_\textnormal{E}}{2} \log (2 \pi e) + \frac{1}{2} \log \left| \textsf{Cov} \{ \mathbb{H}_\textnormal{E}  \mathbf{X}_\textnormal{te} + \mathbf{Z}_\textnormal{E} \}  \right| \qquad \nonumber\\
	&&- \textsf{h}( \mathbf{Z}_\textnormal{E} ) \quad \label{eq: gaussian maximal entropy}\\
	&=& \frac{1}{2} \log \left|\mathbb{H}_\textnormal{E} \cdot \diag{ \mathbf{v} } \cdot \mathbb{H}_\textnormal{E}^\mathsf{T}  + \mathbb{I}_{n_\textnormal{E}}\right|, \label{eq: I(s_te;y_e) upper bound1}
\end{IEEEeqnarray}
where \eqref{eq: gaussian maximal entropy} follows from the principle of maximum entropy, i.e., the entropy of $\bigl( \mathbb{H}_\textnormal{E}  \mathbf{X}_\textnormal{te} + \mathbf{Z}_\textnormal{E}\bigr)$ can be upper-bounded by the entropy of a Gaussian random vector with the same covariance matrix. With \eqref{eq: texp pdf}, we can obtain the vector $\mathbf{v}$ in \eqref{eq: g2 function}.

Substituting \eqref{eq: I(s_te;y_b) lower bound1} and \eqref{eq: I(s_te;y_e) upper bound1} into \eqref{eq: Cs bound2}, we obtain \eqref{eq: Cs lb case i}.

\subsection{Proof of Achievable Secrecy Rate for Case II}\label{sec: Cs lb case ii}

To derive a lower bound on $\textsf{I}( \mathbf{X}_\textnormal{te};\mathbf{Y}_\textnormal{B} )$, we decompose $\mathbb{H}_\textnormal{B}$ via SVD as $\mathbb{H}_\textnormal{B}=\mathbb{U}_\textnormal{B} \mathbb{S}_\textnormal{B} \mathbb{V}_\textnormal{B}$. Here, the orthogonal matrices $\mathbb{U}_\textnormal{B}\in\mathcal{R}^{ n_\textnormal{B} \times n_\textnormal{B} }$, $\mathbb{V}_\textnormal{B}\in \mathcal{R}^{ n_\textnormal{T} \times n_\textnormal{T} } $, and the rectangular diagonal matrix $\mathbb{S}_\textnormal{B} \in \mathcal{R}^{ n_\textnormal{B} \times n_\textnormal{T} } $, with $\mathbb{S}_\textnormal{B}^{n_\textnormal{T}}$ being the first $n_\textnormal{T}$ rows. We bound $\textsf{I}( \mathbf{X}_\textnormal{te};\mathbf{Y}_\textnormal{B} )$ by
\begin{IEEEeqnarray}{rCl}
	\IEEEeqnarraymulticol{3}{l}{%
		\textsf{I}( \mathbf{X}_\textnormal{te};\mathbf{Y}_\textnormal{B} ) 
	}\nonumber\\ \quad
	&=& \textsf{I} \bigl( \mathbf{X}_\textnormal{te}; \mathbb{S}_\textnormal{B} \mathbb{V}_\textnormal{B} \mathbf{X}_\textnormal{te} + \mathbf{Z}_\textnormal{B}^\prime \bigr) \label{eq: svd1} \\
	&=& \textsf{I} \bigl( \mathbf{X}_\textnormal{te}; \mathbb{S}_\textnormal{B}^{n_\textnormal{T}} \mathbb{V}_\textnormal{B} \mathbf{X}_\textnormal{te} + \mathbf{Z}_\textnormal{B}^{\prime\prime} \bigr) \\
	&=& \textsf{I} \left( \mathbf{X}_\textnormal{te}; \mathbf{X}_\textnormal{te} +   \mathbb{V}_\textnormal{B}^\mathsf{T} \bigl(\mathbb{S}_\textnormal{B}^{n_\textnormal{T}}\bigr)^{-1} \mathbf{Z}_\textnormal{B}^{\prime\prime} \right) \\
	&=& \textsf{h} \left(\mathbf{X}_\textnormal{te} +  \mathbb{V}_\textnormal{B}^\mathsf{T} \bigl(\mathbb{S}_\textnormal{B}^{n_\textnormal{T}}\bigr)^{-1} \mathbf{Z}_\textnormal{B}^{\prime\prime} \right) -\textsf{h} \left( \mathbb{V}_\textnormal{B}^\mathsf{T} \bigl(\mathbb{S}_\textnormal{B}^{n_\textnormal{T}}\bigr)^{-1} \mathbf{Z}_\textnormal{B}^{\prime\prime} \right) \label{eq: 23} \quad \\
	&\geq& \frac{n_\textnormal{T}}{2} \log\left( 1 +  \frac{ e^{ \frac{2}{ n_\textnormal{T} } \sum_{i=1}^{n_\textnormal{T}} \textsf{h}( {X}_{\textnormal{te},i} ) } }{ e^{ \frac{2}{ n_\textnormal{T} } \textsf{h} \bigl( \mathbb{V}_\textnormal{B}^\mathsf{T} \left(\mathbb{S}_\textnormal{B}^{n_\textnormal{T}}\right)^{-1} \mathbf{Z}_\textnormal{B}^{\prime\prime} \bigr) } } \right), \label{eq: epi1} 
\end{IEEEeqnarray}
where $\mathbf{Z}_\textnormal{B}^\prime = \mathbb{U}_\textnormal{B}^\mathsf{T} \mathbf{Z}_\textnormal{B}$, $\mathbf{Z}_\textnormal{B}^{\prime\prime}$ represents the first $n_\textnormal{T}$ rows of $\mathbf{Z}_\textnormal{B}^{\prime}$, and \eqref{eq: epi1} holds by the EPI. Note that $\mathbb{V}_\textnormal{B}^\mathsf{T} (\mathbb{S}_\textnormal{B}^{n_\textnormal{T}})^{-1} \mathbf{Z}_\textnormal{B}^{\prime\prime}$ is an $n_\textnormal{T}$-dimensional Gaussian random vector since $\mathbf{Z}_\textnormal{B}^{\prime\prime} \sim \mathcal{N}\left( \mathbf{0}_{ n_\textnormal{T} }, \mathbb{I}_{ n_\textnormal{T} }\right)$. Its covariance matrix is given by
\begin{IEEEeqnarray}{rCl}
	\IEEEeqnarraymulticol{3}{l}{%
		\textsf{Cov} \left\{\mathbb{V}_\textnormal{B}^\mathsf{T} (\mathbb{S}_\textnormal{B}^{n_\textnormal{T}})^{-1} \mathbf{Z}_\textnormal{B}^{\prime\prime} \right\}
	}\nonumber\\ \quad	
	&=& \mathbb{V}_\textnormal{B}^\mathsf{T} \bigl( \mathbb{S}_\textnormal{B}^{n_\textnormal{T}} \bigr)^{-1} 
	\bigl( \mathbb{S}_\textnormal{B}^{n_\textnormal{T}} \bigr)^{-\mathsf{T}} \mathbb{V}_\textnormal{B} \qquad \\
	&=& \left( \mathbb{V}_\textnormal{B}^\mathsf{T} \bigl( \mathbb{S}_\textnormal{B}^{n_\textnormal{T}} \bigr)^\mathsf{T} 
	\bigl( \mathbb{S}_\textnormal{B}^{n_\textnormal{T}} \bigr) \mathbb{V}_\textnormal{B} \right)^{-1} \\
	&=& \left( \mathbb{V}_\textnormal{B}^\mathsf{T} \mathbb{S}_\textnormal{B}^\mathsf{T} \mathbb{S}_\textnormal{B} \mathbb{V}_\textnormal{B} \right)^{-1} \\
	&=& \left(  \mathbb{H}_\textnormal{B}^\mathsf{T} \mathbb{H}_\textnormal{B} \right)^{-1}. \label{eq: cov1}
\end{IEEEeqnarray}
Therefore,
\begin{IEEEeqnarray}{rCl}
	\IEEEeqnarraymulticol{3}{l}{%
		\textsf{h} \left( \mathbb{V}_\textnormal{B}^\mathsf{T} \bigl(\mathbb{S}_\textnormal{B}^{n_\textnormal{T}} \bigr)^{-1} \mathbf{Z}_\textnormal{B}^{\prime\prime} \right)
	}\nonumber\\ \quad	
	&=& \frac{n_\textnormal{T}}{2} \log(2\pi e) + \frac{1}{2} \log \bigl| \textsf{Cov} \left\{\mathbb{V}_\textnormal{B}^\mathsf{T} (\mathbb{S}_\textnormal{B}^{n_\textnormal{T}})^{-1} \mathbf{Z}_\textnormal{B}^{\prime\prime} \right\} \bigr| \quad \\
	& =& \frac{n_\textnormal{T}}{2} \log(2\pi e) - \frac{1}{2} \log \left| \mathbb{H}_\textnormal{B}^\mathsf{T} \mathbb{H}_\textnormal{B} \right|. \label{eq: entropy of gaussian pdf}
\end{IEEEeqnarray}
Substituting \eqref{eq: entropy of gaussian pdf} into \eqref{eq: epi1}, we have
\begin{IEEEeqnarray}{rCl}
	\textsf{I}( \mathbf{X}_\textnormal{te};\mathbf{Y}_\textnormal{B} ) 
	&\geq & \frac{n_\textnormal{T}}{2} \log \left\{ 1 + \left| \diag{\mathbf{p}} \right|^{\frac{1}{ n_\textnormal{T} } } \cdot \left|  \mathbb{H}_\textnormal{B}^\mathsf{T} \mathbb{H}_\textnormal{B}  \right|^{\frac{1}{n_\textnormal{T}}}   \right\}. \label{eq: I(s_te;y_b) lower bound} \qquad
\end{IEEEeqnarray}

To derive an upper bound on $\textsf{I}( \mathbf{X}_\textnormal{te};\mathbf{Y}_\textnormal{E} )$, we also decompose $\mathbb{H}_\textnormal{E}$ via SVD as $\mathbb{H}_\textnormal{E}=\mathbb{U}_\textnormal{E} \mathbb{S}_\textnormal{E} \mathbb{V}_\textnormal{E}$, where the orthogonal matrices $\mathbb{U}_\textnormal{E}\in\mathcal{R}^{ n_\textnormal{E} \times n_\textnormal{E} }$, $\mathbb{V}_\textnormal{E}\in \mathcal{R}^{ n_\textnormal{T} \times n_\textnormal{T} } $, and the rectangular diagonal matrix $\mathbb{S}_\textnormal{E} \in \mathcal{R}^{ n_\textnormal{E} \times n_\textnormal{T} } $, with $\mathbb{S}_\textnormal{E}^{n_\textnormal{T}}$ being the first $n_\textnormal{T}$ rows. Similarly, we derive
\begin{IEEEeqnarray}{rCl}
	\IEEEeqnarraymulticol{3}{l}{%
		\textsf{I}( \mathbf{X}_\textnormal{te};\mathbf{Y}_\textnormal{E} ) 
	}\nonumber\\ 
	&=& \textsf{h} \left( \mathbf{X}_\textnormal{te} +  \mathbb{V}_\textnormal{E}^\mathsf{T} \bigl( \mathbb{S}_\textnormal{E}^{n_\textnormal{T}} \bigr)^{-1} \mathbf{Z}_\textnormal{E}^{\prime\prime} \right) -\textsf{h} \left( \mathbb{V}_\textnormal{E}^\mathsf{T} 
	\bigl( \mathbb{S}_\textnormal{E}^{n_\textnormal{T}} \bigr)^{-1} \mathbf{Z}_\textnormal{E}^{\prime\prime} \right) \label{eq: 21} \quad \\
	&=& \textsf{h} \left( \mathbf{X}_\textnormal{te} +  \mathbb{V}_\textnormal{E}^\mathsf{T} \bigl( \mathbb{S}_\textnormal{E}^{n_\textnormal{T}} \bigr)^{-1} \mathbf{Z}_\textnormal{E}^{\prime\prime} \right) - \frac{n_\textnormal{T}}{2} \log(2\pi e) \nonumber\\
	&&  + \frac{1}{2} \log \left|  \mathbb{H}_\textnormal{E}^\mathsf{T} \mathbb{H}_\textnormal{E}  \right| \label{eq: 31}\\
	&\leq& \frac{1}{2} \log \left| \textsf{Cov} \bigl\{ \mathbf{X}_\textnormal{te} +  \mathbb{V}_\textnormal{E}^\mathsf{T} \bigl( \mathbb{S}_\textnormal{E}^{n_\textnormal{T}} \bigr)^{-1} \mathbf{Z}_\textnormal{E}^{\prime\prime} \bigr\}  \right| + \frac{1}{2} \log \left|  \mathbb{H}_\textnormal{E}^\mathsf{T} \mathbb{H}_\textnormal{E}  \right| \label{eq: 43} \quad \\
	&=& \frac{1}{2} \log \left| \diag{\mathbf{v}} \cdot \mathbb{H}_\textnormal{E}^\mathsf{T} \mathbb{H}_\textnormal{E} +  \mathbb{I}_{ n_\textnormal{T} } \right|, \label{eq: 43-2}
\end{IEEEeqnarray}
where $\mathbf{Z}_\textnormal{E}^{\prime\prime}$ denotes the first $n_\textnormal{T}$ rows of $\mathbb{U}_\textnormal{E}^\mathsf{T} \mathbf{Z}_\textnormal{E}$, \eqref{eq: 31} follows because $\mathbb{V}_\textnormal{E}^\mathsf{T} (\mathbb{S}_\textnormal{E}^{n_\textnormal{T}})^{-1} \mathbf{Z}_\textnormal{E}^{\prime\prime} \sim \mathcal{N} \left( \mathbf{0}_{ n_\textnormal{T} }, \left(  \mathbb{H}_\textnormal{E}^\mathsf{T} \mathbb{H}_\textnormal{E} \right)^{-1} \right)$, and \eqref{eq: 43} holds by the principle of maximum entropy.

Substituting \eqref{eq: I(s_te;y_b) lower bound} and \eqref{eq: 43-2} into \eqref{eq: Cs bound2} yields \eqref{eq: Cs lb case ii}.

\section{MIMO-VLC Wiretap Channel Beamforming Design}\label{sec: beamforming design}
This section proposes beamforming schemes to enhance the secrecy performance of the MIMO-VLC wiretap channels. We first analyze Case I, and then proceed to Case II.

\vspace{-2mm}
\subsection{Beamforming Design for Case I}
\subsubsection{Fully-connected Beamforming Design}\label{sec: full-connected Case I}
A linear transformation is applied to the input signals across different LEDs such that
\begin{IEEEeqnarray}{rCl}
	\subnumberinglabel{channel model beamforming}
	\mathbf{Y}_\textnormal{B} &=& \mathbb{H}_\textnormal{B} ( \mathbb{W}^\mathsf{T} \mathbf{X} + \amp \mathbf{d} ) + \mathbf{Z}_\textnormal{B},\\
	\mathbf{Y}_\textnormal{E} &=& \mathbb{H}_\textnormal{E} ( \mathbb{W}^\mathsf{T} \mathbf{X} + \amp \mathbf{d} ) + \mathbf{Z}_\textnormal{E}, 
\end{IEEEeqnarray}
where $\mathbb{W} \in \mathcal{R}^{ n_\textnormal{T} \times n_\textnormal{T}} $ is the beamforming matrix and $\mathbf{d} \in \mathcal{R}^{ n_\textnormal{T}} $ is a bias vector. Note that \eqref{channel model} represents a special case of \eqref{channel model beamforming} when $\mathbb{W}=\mathbb{I}_{n_\textnormal{T}}$ and $\mathbf{d}=\mathbf{0}_{n_\textnormal{T}}$. In \eqref{channel model beamforming}, each input $X_j$ is fed to all LEDs through the beamformer $\mathbb{W}$, while in \eqref{channel model}, $X_j$ is directly connected solely to the $j$-th LED, $j\in\{1,\cdots,n_\textnormal{T}\}$. For clarity, we refer to \eqref{channel model beamforming} as the fully-connected beamforming scheme and \eqref{channel model} as the direct-connected beamforming scheme, and their comparison is illustrated in Fig.~\ref{fig2}. 

\begin{figure}[ht]
	\centering
	\subfloat[Direct-connected beamformer.]{
		\includegraphics[width=1.67in]{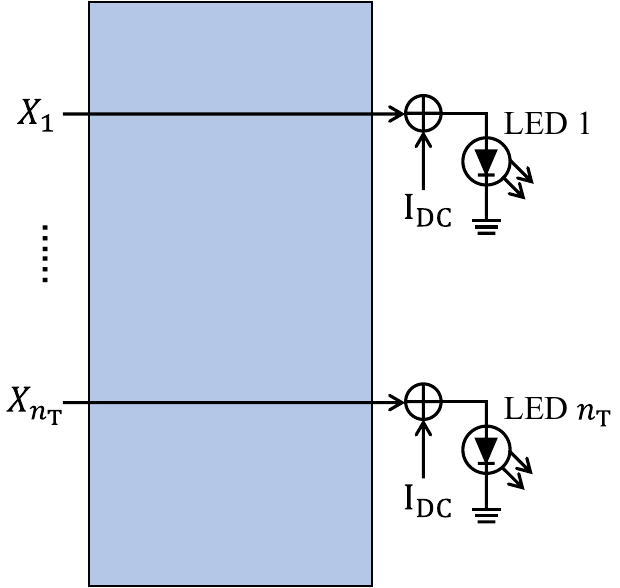}
	}\hspace{-3mm}	
	\subfloat[Fully-connected beamformer.]{
		\includegraphics[width=1.67in]{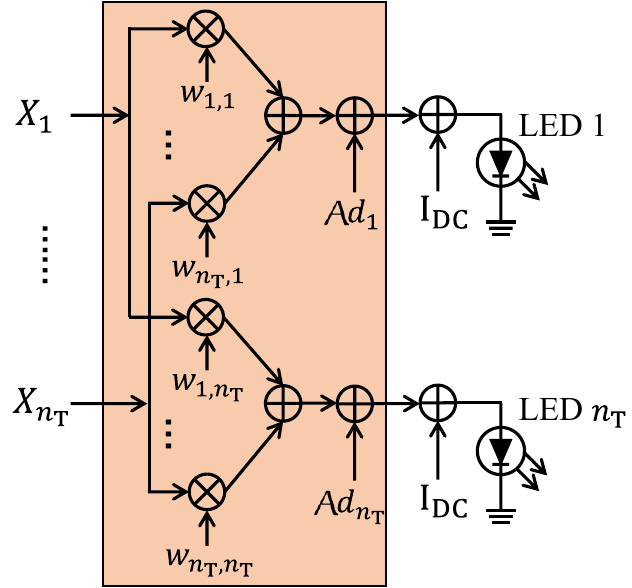}
	}
	\caption{Architectures of two beamforming schemes: (a) Direct-connected beamformer: the $j$-th LED is only connected to the $j$-th input $X_j$; (b) Fully-connected beamformer: each LED is connected to all inputs $X_j$s, $j\in\{1,\cdots,n_\textnormal{T}\}$.}
	\label{fig2}
\end{figure}

Denote $\mathbb{W} = (\mathbf{w}_1,\cdots,\mathbf{w}_{n_\textnormal{T}})$. To satisfy the peak-intensity constraint for LEDs, it suffices that $\forall j\in\{1,\cdots,n_\textnormal{T}\}$,
\begin{IEEEeqnarray}{rCl}
	\subnumberinglabel{eq: wi}
	\| \mathbf{w}_j \|_1 + d_j &\leq& 1,\\
	-\| \mathbf{w}_j \|_1 + d_j &\geq& -1.
\end{IEEEeqnarray}
Additionally, satisfying the average-intensity constraint requires that $\forall j\in\{1,\cdots,n_\textnormal{T}\}$,
\begin{IEEEeqnarray}{rCl}
	\mathbf{w}_j^\mathsf{T} \bm{\beta} + \frac{1}{2} d_j = \beta_j. \label{eq: optimal d}
\end{IEEEeqnarray}
We have $\| \mathbf{w}_j \|_1 \leq 1$ from \eqref{eq: wi}. Combining this with \eqref{eq: optimal d}, a feasible $d_j$ exists if $|d_j|\leq 1 - \| \mathbf{w}_j \|_1 $. Hence, both constraints are satisfied if and only if $\forall j\in \{1,\cdots,n_\textnormal{T}\}$,
\begin{IEEEeqnarray}{rCl}
	\subnumberinglabel{eq: wi and d}
	\| \mathbf{w}_j \|_1 &\leq& 1,\\
	|\mathbf{w}_j^\mathsf{T} \bm{\beta}-\beta_j| &\leq& \frac{1}{2} - \frac{1}{2} \| \mathbf{w}_j \|_1.
\end{IEEEeqnarray}
Note that the equivalent channel matrices of Bob and Eve become $\mathbb{H}_\textnormal{B}\mathbb{W}^\mathsf{T}$ and  $\mathbb{H}_\textnormal{E}\mathbb{W}^\mathsf{T}$, respectively\footnote{It should be noted that constants $\amp\mathbb{H}_\textnormal{B}\mathbf{d}$ and $\amp\mathbb{H}_\textnormal{E}\mathbf{d}$ only have an effect on the input constraint without influencing the value of achievable secrecy rate.}. Substituting them into \eqref{eq: Cs lb case i}, the optimal fully-connected beamformer $\mathbb{W}^\star_\textnormal{FC}$ maximizing the achievable secrecy rate can be obtained by:
\begin{align}
	\mathcal{P}_1:
	 \mathbb{W}_\textnormal{FC}^\star = &\argmax_{\mathbb{W}} \bigl\{ \mathsf{R}_\textnormal{s}^\textnormal{I}(\mathbb{H}_\textnormal{B}\mathbb{W}^\mathsf{T},\mathbb{H}_\textnormal{E}\mathbb{W}^\mathsf{T}) \bigr\}  
	 \\ 
	& \mathrm{s.t.} \biggl\{ \begin{array}{l}
		\| \mathbf{w}_j \|_1 \leq 1, \forall j,\\
		|\mathbf{w}_j^\mathsf{T} \bm{\beta} - \beta_{j} | \leq \frac{1}{2} - \frac{1}{2} \| \mathbf{w}_j \|_1, \forall j. 
	\end{array}\Bigr. \label{eq: w sons}
\end{align}

Next, we analyze the fully-connected ZF beamforming scheme for the MIMO-VLC wiretap channel. This scheme restricts the beamformer to lie within Eve's nullspace, i.e.,
\begin{IEEEeqnarray}{rCl}
	\mathbb{H}_\textnormal{E} \mathbb{W}^\mathsf{T} = \mathbf{0}_{ n_\textnormal{E} \times n_\textnormal{T} }. \label{eq: zf}
\end{IEEEeqnarray}
Consequently, the optimal fully-connected ZF beamformer $\mathbb{W}^\star_{\textnormal{ZF}}$ is obtained by:
\begin{IEEEeqnarray}{rCl}
	\mathcal{P}_2:
	\mathbb{W}^\star_{\textnormal{ZF}} & =& \argmax_{\mathbb{W}} \bigl\{ \mathsf{R}_\textnormal{s}^\textnormal{I}(\mathbb{H}_\textnormal{B}\mathbb{W}^\mathsf{T},\mathbb{H}_\textnormal{E}\mathbb{W}^\mathsf{T}) \bigr\} \\
	&& \mathrm{s.t.} \Biggl\{ \begin{array}{l}
	\| \mathbf{w}_j \|_1 \leq 1, \forall j, \\
	|\mathbf{w}_j^\mathsf{T} \bm{\beta} - \beta_{j} | \leq \frac{1}{2} - \frac{1}{2} \| \mathbf{w}_j \|_1, \forall j,\\
	\mathbb{H}_\textnormal{E} \mathbb{W}^\mathsf{T} = \mathbf{0}_{ n_\textnormal{E} \times n_\textnormal{T} }.
	\end{array}\Bigr.  \label{eq: W2}
\end{IEEEeqnarray}

\subsubsection{Sub-connected Beamforming Design}
This section proposes a low-complexity beamforming scheme as an alternative
to the fully-connected architecture. Denote $\mathcal{I}$ as a $\textnormal{rank}(\mathbb{H}_\textnormal{B})$-size subset of $\{1,2,\cdots,n_\textnormal{T}\}$ and it is chosen such that $|\mathbb{H}_{\textnormal{B},\mathcal{I}}|\neq 0$, where $\mathbb{H}_{\textnormal{B},\mathcal{I}}$ is the submatrix of $\mathbb{H}_\textnormal{B}$ with columns indexed by $\mathcal{I}$. For convenience, $\mathcal{I}_\textnormal{all}$ denotes the set of all $\mathcal{I}$s satisfying $|\mathbb{H}_{\textnormal{B},\mathcal{I}}|\neq 0$.\footnote{For example, if $\mathbb{H}_\textnormal{B}$ is a $(2\times 3)$-dimensional matrix with $\mathbb{H}_\textnormal{B}=(\mathbf{h}_1,\mathbf{h}_2,\mathbf{h}_3)$ and $\textnormal{rank}(\mathbb{H}_\textnormal{B})=2$. Then $\mathcal{I}$ may be $\{1,2\}$, $\{1,3\}$, $\{2,3\}$, the corresponding $\mathbb{H}_{\textnormal{B},\mathcal{I}}$ is $(\mathbf{h}_1,\mathbf{h}_2)$, $(\mathbf{h}_1,\mathbf{h}_3)$, $(\mathbf{h}_2,\mathbf{h}_3)$, and $\mathcal{I}_\textnormal{all}$ is $\{\{1,2\},\{1,3\}, \{2,3\}\}$.} Using $\mathcal{I}$, the channel outputs in \eqref{channel model} can be reformulated as
\begin{IEEEeqnarray}{rCl}
	\subnumberinglabel{eq: channel model 2}
	\mathbf{Y}_\textnormal{B} 
	&=& \mathbb{H}_{ \textnormal{B}, \mathcal{I} } \mathbf{X}_\mathcal{I} + \mathbb{H}_{ \textnormal{B}, \mathcal{I}^\textnormal{c} } \mathbf{X}_{\mathcal{I}^\textnormal{c}}
	+ \mathbf{Z}_\textnormal{B},\\
	\mathbf{Y}_\textnormal{E} 
	&=& \mathbb{H}_{ \textnormal{E}, \mathcal{I} } \mathbf{X}_\mathcal{I} + \mathbb{H}_{ \textnormal{E}, \mathcal{I}^\textnormal{c} } \mathbf{X}_{\mathcal{I}^\textnormal{c}}
	+ \mathbf{Z}_\textnormal{E},
\end{IEEEeqnarray}
where $\mathcal{I}^\textnormal{c}$ denotes the complement of $\mathcal{I}$, i.e., $\mathcal{I}^\textnormal{c} = \{1,2,\cdots,n_\textnormal{T}\}\backslash\mathcal{I}$;
$\mathbf{X}_\mathcal{I}$ and $\mathbf{X}_{\mathcal{I}^\textnormal{c}}$ are the subvectors of $\mathbf{X}$ with entries indexed by $\mathcal{I}$ and $\mathcal{I}^\textnormal{c}$, respectively\footnote{For example, if $\mathbf{X}=(8,6,7)^\mathsf{T}$, $\mathcal{I}=\{1,2\}$, and $\mathcal{I}^\textnormal{c}=\{3\}$, then $\mathbf{X}_\mathcal{I}=(8,6)^\mathsf{T}$, and $\mathbf{X}_{\mathcal{I}^\textnormal{c}}=7$.}; $\mathbb{H}_{ \textnormal{B}, \mathcal{I}^\textnormal{c} }$, $\mathbb{H}_{ \textnormal{E}, \mathcal{I} }$, and $\mathbb{H}_{ \textnormal{E}, \mathcal{I}^\textnormal{c} }$ are the submatrices of $\mathbb{H}_\textnormal{B}$, $\mathbb{H}_\textnormal{E}$ and $\mathbb{H}_\textnormal{E}$ with columns indexed by $\mathcal{I}^\textnormal{c}$, $\mathcal{I}$, and $\mathcal{I}^\textnormal{c}$, respectively. 

Let $\mathbb{B}$ be an $n_\textnormal{B} \times (n_\textnormal{T}-n_\textnormal{B})$-dimensional matrix with $\mathbb{B}=(\mathbf{b}_1,\cdots,\mathbf{b}_{n_\textnormal{T}-n_\textnormal{B}})$, and $\mathbf{c}$ be an $( n_\textnormal{T}-n_\textnormal{B} )$-dimensional vector with $\mathbf{c}=(c_1,\cdots,c_{n_\textnormal{T}-n_\textnormal{B}})^\mathsf{T}$. The sub-connected beamforming scheme designs the subvector $\mathbf{X}_{\mathcal{I}^\textnormal{c}}$ to be a linear combination of the subvector $\mathbf{X}_\mathcal{I}$ such that
\begin{IEEEeqnarray}{rCl}
	\mathbf{X}_{\mathcal{I}^\textnormal{c}} = \mathbb{B}^\mathsf{T} \mathbf{X}_\mathcal{I} + \amp \mathbf{c} . \label{eq: liear beamforming}
\end{IEEEeqnarray}
Substituting it into \eqref{eq: channel model 2}, we obtain
\begin{IEEEeqnarray}{rCl}
	\subnumberinglabel{eq: channel model sub-connect}
	\mathbf{Y}_\textnormal{B} 
	&=& (\mathbb{H}_{ \textnormal{B}, \mathcal{I} } 
	+ \mathbb{H}_{ \textnormal{B}, \mathcal{I}^\textnormal{c} } \mathbb{B}^\mathsf{T}) \mathbf{X}_\mathcal{I} 
	+ \mathbf{Z}_\textnormal{B} + \amp \mathbb{H}_{ \textnormal{B}, \mathcal{I}^\textnormal{c} } \mathbf{c},\\
	\mathbf{Y}_\textnormal{E} 
	&=& (\mathbb{H}_{ \textnormal{E}, \mathcal{I} } 
	+ \mathbb{H}_{ \textnormal{E}, \mathcal{I}^\textnormal{c} } \mathbb{B}^\mathsf{T}) \mathbf{X}_\mathcal{I} 
	+ \mathbf{Z}_\textnormal{E} + \amp \mathbb{H}_{ \textnormal{E}, \mathcal{I}^\textnormal{c} } \mathbf{c}.
\end{IEEEeqnarray}
Fig.~\ref{fig3} illustrates this sub-connected beamforming architecture. Note that \eqref{eq: channel model sub-connect} is a special case of the fully-connected beamforming scheme \eqref{channel model beamforming}. For example, when $\mathcal{I}=\{1,\cdots,n_\textnormal{B}\}$ and $|\mathbb{H}_{ \textnormal{B}, \mathcal{I} }|\neq 0$, define 
\begin{IEEEeqnarray}{rCl}
	\mathbb{W} &=& \Biggl( \begin{array}{ll}
		\mathbb{I}_{n_\textnormal{B}} & \mathbb{B}\\
		\mathbf{0}_{(n_\textnormal{T}-n_\textnormal{B}) \times n_\textnormal{B}} & \mathbf{0}_{(n_\textnormal{T}-n_\textnormal{B}) \times (n_\textnormal{T}-n_\textnormal{B}) }
	\end{array} \Biggr),\\
	\mathbf{d} &=& \Bigl(\begin{array}{l}
		\mathbf{0}_{n_\textnormal{B} }\\
		\mathbf{c}
	\end{array}\Bigr).
\end{IEEEeqnarray}
Substituting them into \eqref{channel model beamforming} yields \eqref{eq: channel model sub-connect}. 

\begin{figure}[t]
	\centering	
	\includegraphics[width=2.5in]{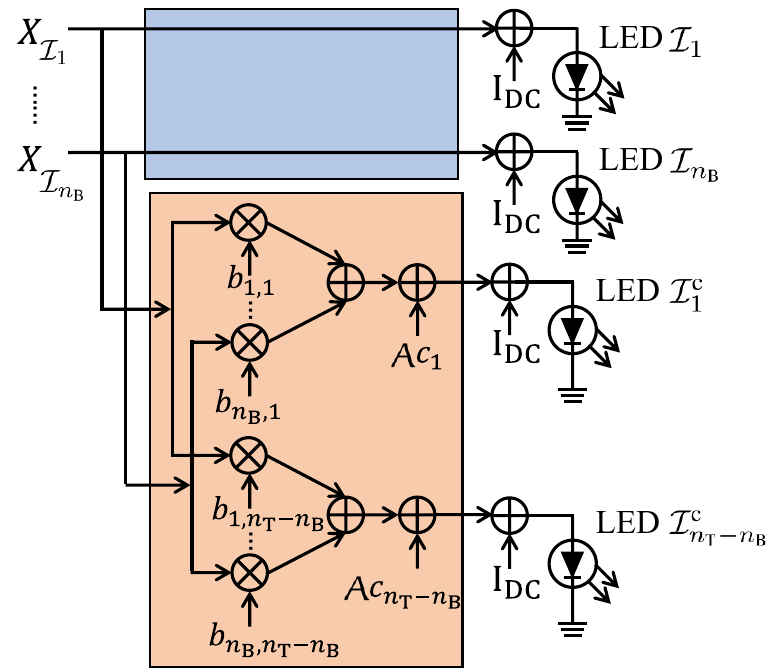}
	\caption{Architecture of sub-connected beamforming scheme: the $\mathcal{I}_j$-th LED is only connected to the $\mathcal{I}_j$-th input $X_{\mathcal{I}_j}$, while the $\mathcal{I}^\textnormal{c}_k$-th LED is connected to all inputs $X_{\mathcal{I}_j}$s, where $\mathcal{I}_j\in\mathcal{I}$, $j\in \{ 1,\cdots,n_\textnormal{B} \}$, $\mathcal{I}^\textnormal{c}_k\in\mathcal{I}^\textnormal{c}$, and $k\in \{ 1,\cdots,n_\textnormal{T}-n_\textnormal{B} \}$.}
	\label{fig3}
\end{figure}

Under the sub-connected beamforming architecture, the peak- and average-intensity constraints are satisfied if $\forall j \in \{ 1,\cdots,n_\textnormal{T} -n_\textnormal{B} \}$,
\begin{IEEEeqnarray}{rCl}
	\subnumberinglabel{eq: bj}
	-\| \mathbf{b}_j \|_1 + c_j &\geq& -1, \\
	\| \mathbf{b}_j \|_1 + c_j &\leq& 1,\\
	\mathbf{b}_j^\mathsf{T} \bm{\beta}_{\mathcal{I}} + \frac{1}{2}c_j & = & \beta_{\mathcal{I}^\textnormal{c}_j}, \label{eq: 51c}
\end{IEEEeqnarray}
where $\bm{\beta}_{\mathcal{I}}$ and $\bm{\beta}_{\mathcal{I}^\textnormal{c}}$ denote the subvectors of $\bm{\beta}$ with entries indexed by $\mathcal{I}$ and $\mathcal{I}^\textnormal{c}$, respectively, and $\beta_{\mathcal{I}^\textnormal{c}_j}$ is the $j$-th entry of $\bm{\beta}_{\mathcal{I}^\textnormal{c}}$. Analogous to \eqref{eq: wi and d}, the inequalities in \eqref{eq: bj} are equivalent to
\begin{IEEEeqnarray}{rCl}
	\| \mathbf{b}_j \|_1 &\leq& 1,\\
	|\mathbf{b}_j^\mathsf{T} \bm{\beta}_{\mathcal{I}} - \beta_{\mathcal{I}^\textnormal{c}_j} | &\leq& \frac{1}{2} - \frac{1}{2} \| \mathbf{b}_j \|_1.
\end{IEEEeqnarray}
Note that in \eqref{eq: channel model sub-connect}, the equivalent channel matrices of Bob and Eve are $(\mathbb{H}_{ \textnormal{B}, \mathcal{I} }+\mathbb{H}_{ \textnormal{B}, \mathcal{I}^\textnormal{c} } \mathbb{B}^\mathsf{T})$ and  $(\mathbb{H}_{ \textnormal{E},\mathcal{I} }+\mathbb{H}_{ \textnormal{E}, \mathcal{I}^\textnormal{c} } \mathbb{B}^\mathsf{T})$, respectively. Substituting them into \eqref{eq: Cs lb case i}, the optimal sub-connected beamformer $\mathbb{B}_\textnormal{SC}^\star$ maximizing the achievable secrecy rate is found by:
\begin{IEEEeqnarray}{rCl}
	\IEEEeqnarraymulticol{3}{l}{%
		\mathcal{P}_3:\ (\mathbb{B}_\textnormal{SC}^\star,\mathcal{I}_\textnormal{SC}^\star) = 
	}\nonumber\\ \qquad\
	&&\argmax_{\mathbb{B},\mathcal{I}} \bigl\{ \mathsf{R}_\textnormal{s}^\textnormal{I}(\mathbb{H}_{ \textnormal{B}, \mathcal{I} }+\mathbb{H}_{ \textnormal{B}, \mathcal{I}^\textnormal{c} } \mathbb{B}^\mathsf{T},\mathbb{H}_{ \textnormal{E}, \mathcal{I} }+\mathbb{H}_{ \textnormal{E}, \mathcal{I}^\textnormal{c} } \mathbb{B}^\mathsf{T} ) \bigr\} \qquad\\
	&& \mathrm{s.t.} \biggl\{ \begin{array}{l}
		\| \mathbf{b}_j \|_1 \leq 1, \forall j ,\\
		|\mathbf{b}_j^\mathsf{T} \bm{\beta}_\mathcal{I} - \beta_{\mathcal{I}^\textnormal{c}_j} | \leq \frac{1}{2} - \frac{1}{2} \| \mathbf{b}_j \|_1,\forall j .	
	\end{array}\Bigr.  \label{eq: b cons}
\end{IEEEeqnarray}

Next, we analyze the sub-connected ZF beamforming scheme for the MIMO-VLC wiretap channel. This is achieved by imposing the constraint:
\begin{IEEEeqnarray}{rCl}
	\mathbb{H}_{ \textnormal{E}, \mathcal{I}^\textnormal{c} } \mathbb{B}^\mathsf{T} = - \mathbb{H}_{ \textnormal{E}, \mathcal{I} }. \label{eq: zf2}
\end{IEEEeqnarray}
Thus, the optimal sub-connected ZF beamformer $\mathbb{B}_\textnormal{ZF}^\star$ is found by:
{
\begin{IEEEeqnarray}{rCl}
	\IEEEeqnarraymulticol{3}{l}{%
		\mathcal{P}_4:\ (\mathbb{B}_\textnormal{ZF}^\star,\mathcal{I}_\textnormal{ZF}^\star) = 
	}\nonumber\\ \qquad\
	&&\argmax_{\mathbb{B},\mathcal{I}} \bigl\{ \mathsf{R}_\textnormal{s}^\textnormal{I}(\mathbb{H}_{ \textnormal{B}, \mathcal{I} }+\mathbb{H}_{ \textnormal{B}, \mathcal{I}^\textnormal{c} } \mathbb{B}^\mathsf{T},\mathbb{H}_{ \textnormal{E}, \mathcal{I} }+\mathbb{H}_{ \textnormal{E}, \mathcal{I}^\textnormal{c} } \mathbb{B}^\mathsf{T} ) \bigr\}\qquad\\
	&& \mathrm{s.t.} \Biggl\{ \begin{array}{l}
		\| \mathbf{b}_j \|_1 \leq 1, \forall j,\\
		|\mathbf{b}_j^\mathsf{T} \bm{\beta}_{\mathcal{I}} - \beta_{\mathcal{I}^\textnormal{c}_j} | \leq \frac{1}{2} - \frac{1}{2} \| \mathbf{b}_j \|_1,\forall j,\\
		\mathbb{H}_{ \textnormal{E}, \mathcal{I}^\textnormal{c} } \mathbb{B}^\mathsf{T} = - \mathbb{H}_{ \textnormal{E}, \mathcal{I} }.	
	\end{array}\Biggr.  \label{eq:45}
\end{IEEEeqnarray}}

However, $\mathcal{P}_4$ may be infeasible. For example, if $\mathbb{H}_{ \textnormal{E}, \mathcal{I}^\textnormal{c} }$ is full column-rank, we set $\mathbb{B} =  ( - \mathbb{H}_{ \textnormal{E}, \mathcal{I}^\textnormal{c} }^{\dagger} \mathbb{H}_{ \textnormal{E}, \mathcal{I} }  )^\mathsf{T}$ to satisfy \eqref{eq: zf2}, where $\mathbb{H}_{ \textnormal{E}, \mathcal{I}^\textnormal{c} }^\dagger = (\mathbb{H}_{ \textnormal{E}, \mathcal{I}^\textnormal{c} }^{\textsf{T}} \mathbb{H}_{ \textnormal{E}, \mathcal{I}^\textnormal{c} } )^{-1} \mathbb{H}_{ \textnormal{E}, \mathcal{I}^\textnormal{c} }^{\textsf{T}}$, is the Moore-Penrose inverse of $\mathbb{H}_{ \textnormal{E}, \mathcal{I}^\textnormal{c} }$. If this $\mathbb{B}$ violates the first two conditions in \eqref{eq:45}, i.e., the peak- and average-intensity constraints, then $\mathcal{P}_4$ has no feasible solution. To address this, we relax \eqref{eq: zf2} and seek an approximate solution $\mathbb{B}_\textnormal{MLSE}^\star$ minimizing the least square error (MLSE) while satisfying the constraints:
\begin{IEEEeqnarray}{rCl}
	\IEEEeqnarraymulticol{3}{l}{%
		\mathcal{P}_5:(\mathbb{B}_\textnormal{MLSE}^\star,\mathcal{I}_\textnormal{MLSE}^\star) =
	}\nonumber\\ \qquad\qquad\qquad	
	&& \argmin\limits_{\mathbb{B},\mathcal{I}} \| \mathbb{H}_{ \textnormal{E}, \mathcal{I}^\textnormal{c} } \mathbb{B}^\mathsf{T} + \mathbb{H}_{ \textnormal{E}, \mathcal{I} } \|_\mathsf{F} \label{eq: B_MLSE}\\
	&& \mathrm{s.t.} \biggl\{ \begin{array}{l}
		\| \mathbf{b}_j \|_1 \leq 1, \forall j ,\\
		|\mathbf{b}_j^\mathsf{T} \bm{\beta}_\mathcal{I} - \beta_{\mathcal{I}^\textnormal{c}_j} | \leq \frac{1}{2} - \frac{1}{2} \| \mathbf{b}_j \|_1,\forall j .		
	\end{array}\Bigr.\quad
\end{IEEEeqnarray}
\subsection{Beamforming Design for Case II}\label{sec: beamformer i}
This section addresses Case II using analysis similar to Sec.~\ref{sec: full-connected Case I}. The fully-connected beamforming scheme for Case II remains identical to \eqref{channel model beamforming} and Fig. \ref{fig2}(b). Using \eqref{eq: Cs lb case ii}, the optimal fully-connected beamformer $\mathbb{W}_\textnormal{FC}^\star$ maximizing the achievable secrecy rate is obtained by:
\begin{IEEEeqnarray}{rCl}
	\mathcal{P}_6: \mathbb{W}_\textnormal{FC}^\star
	& = & \argmax_{\mathbb{W}} \bigl\{ \mathsf{R}_\textnormal{s}^\textnormal{II}(\mathbb{H}_\textnormal{B}\mathbb{W}^\mathsf{T},\mathbb{H}_\textnormal{E}\mathbb{W}^\mathsf{T} \bigr\} \\ 
	&& \mathrm{s.t.} \biggl\{ \begin{array}{l}
	\| \mathbf{w}_j \|_1 \leq 1, \forall j,\\
		|\mathbf{w}_j^\mathsf{T} \bm{\beta} - \beta_{j} | \leq \frac{1}{2} - \frac{1}{2} \| \mathbf{w}_j \|_1,\forall j .
		\end{array}\Bigr.  \label{eq: eq: w cons 3}
\end{IEEEeqnarray}

Note that under Case II, $\mathbb{H}_\textnormal{E}$ is a full column-rank matrix. The fully-connected ZF beamformer satisfying \eqref{eq: zf} {corresponds to} the zero matrix, rendering the ZF-based beamforming scheme infeasible. Furthermore, we observe that the sub-connected beamforming scheme degenerates to the direct-connected beamforming scheme. Specifically, with the index set $\mathcal{I}=\{1,2,\cdots,n_\textnormal{T}\}$, we obtain $\mathbb{H}_\textnormal{B}=\mathbb{H}_{\textnormal{B},\mathcal{I}} $. This configuration directly connects the $j$-th input $X_j$ to the $j$ LED for all $j\in\{1,\cdots,n_\textnormal{T}\}$, which corresponds precisely to the direct-connected beamforming scheme.

\section{Secrecy Rate Maximization Algorithm}\label{sec: algorithm}
This section develops efficient algorithms for solving the beamforming optimization problems $\mathcal{P}_1$-$\mathcal{P}_6$. Note that $\mathcal{P}_1$-$\mathcal{P}_6$ (except $\mathcal{P}_5$) are non-convex due to non-concave objective functions. To address this challenge, we first decompose each problem's objective function into two components corresponding to the mutual information of Bob and Eve, respectively. For each component, we construct a strongly convex surrogate function: Bob's surrogate is derived from a second-order Taylor expansion with a positive semi-definite Hessian, while Eve's from a first-order Taylor expansion of a log-det function. These surrogates are then optimized within a successive convex approximation (SCA) framework, which has been proven to be effective in wireless communications, signal processing, and machine learning \cite{Razaviyayn2014,Scutari2014,Sun2016}. We begin by introducing SCA fundamentals before presenting SCA-based algorithms for the MIMO-VLC wiretap channel beamforming optimization.



\vspace{-2.5mm}
\subsection{Preliminary}
Consider a non-convex problem:
\begin{IEEEeqnarray}{rCl}
\mathcal{P}: \ \min_{\mathbf{x}\in\mathcal{X}} f(\mathbf{x}), \label{eq: non-convex pro}
\end{IEEEeqnarray}
where $f(\mathbf{x})$ is non-convex with respect to $\mathbf{x}$. For simplicity, we assume $\mathcal{X}$ is a convex constraint. To solve such a non-convex problem, SCA first constructs a locally surrogate function to approximate $f(\mathbf{x})$ and then solves the difficult non-convex problem $\mathcal{P}$ via successively solving a series of convex subproblems. Specifically, at $k$-th iteration, the subproblem for $\mathcal{P}$ is given by
\begin{IEEEeqnarray}{rCl}
	\mathcal{P}^\prime:\
	&& \hat{\mathbf{x}}^{(k+1)} =\argmin_{\mathbf{x}\in\mathcal{X}} \widetilde{f}(\mathbf{x};\mathbf{x}^{(k)}), 
\end{IEEEeqnarray}
where $\mathbf{x}^{(k)}$ denotes the current iteration variable, and $\widetilde{f}(\mathbf{x};\mathbf{x}^{(k)})$ denotes the surrogate function of $f(\mathbf{x})$ at $\mathbf{x}^{(k)}$. The surrogate function $\widetilde{f}(\mathbf{x};\mathbf{x}^{(k)})$ is required to satisfy the following conditions:
\begin{itemize}
\item $\widetilde{f}(\mathbf{x}^{(k)};\mathbf{x}^{(k)}) = f(\mathbf{x}^{(k)})$;
\item $\nabla \widetilde{f}(\mathbf{x}^{(k)};\mathbf{x}^{(k)})= \nabla f(\mathbf{x}^{(k)})$;
\item $\widetilde{f}(\mathbf{x};\mathbf{x}^{(k)}) $ is strongly convex on $\mathcal{X}$.
\end{itemize}
The next iteration variable $\mathbf{x}^{(k+1)}$ is generated through
\begin{IEEEeqnarray}{rCl}
\mathbf{x}^{(k+1)} = \mathbf{x}^{(k)} + \gamma^{(k)} (\hat{\mathbf{x}}^{(k+1)}-\mathbf{x}^{(k)}),\label{eq: xk update} 
\end{IEEEeqnarray}
where $\gamma^{(k)}$ is the step size at the $k$-th iteration. The convergence of SCA algorithm is guaranteed by $\lim_{k\rightarrow +\infty} \gamma^{(k)} = 0$ and $\sum_{k=0}^{+\infty} \gamma^{(k)} = +\infty$ \cite{Scutari2014,Sun2016}.

The key to the success of SCA lies in constructing the surrogate function at each iteration. In the following, we separately construct the surrogate functions for problems $\mathcal{P}_1$-$\mathcal{P}_6$ (except $\mathcal{P}_5$) and propose the SCA-aided algorithms to solve them.
\vspace{-3mm}
\subsection{Algorithm for Case I}
\subsubsection{Fully-connected Beamforming Optimization Algorithm} \label{full-connect algorithm}
Note that the objective functions of problems $\mathcal{P}_1$ and $\mathcal{P}_2$ are the same, while their corresponding feasible domains are different. To be convenient, we denote the objective function of problem $\mathcal{P}_1$ or $\mathcal{P}_2$ by $-f(\mathbb{W})$, i.e.,
\begin{align}
	f(\mathbb{W}) = - \mathsf{R}_\textnormal{s}^\textnormal{I}(\mathbb{H}_\textnormal{B}\mathbb{W}^\mathsf{T},\mathbb{H}_\textnormal{E}\mathbb{W}^\mathsf{T}).
\end{align}
Denote the feasible domain of problem $\mathcal{P}_1$ by $\mathcal{W}_1$, and the feasible domain of problem $\mathcal{P}_2$ by $\mathcal{W}_2$. Without ambiguity, the following algorithm applies to both $\mathcal{P}_1$ and $\mathcal{P}_2$, and the difference is in the feasible domains. 

To begin with, we divide $f(\mathbb{W}) $ into two parts such that $f(\mathbb{W})  = f_\textnormal{B}(\mathbb{W}) + f_\textnormal{E}(\mathbb{W}) $ with
\begin{small}
\begin{IEEEeqnarray}{rCl}
f_\textnormal{B}(\mathbb{W}) 
&=& - \frac{n_\textnormal{B}}{2} \log \left( 1 + \left| \mathbb{H}_\textnormal{B}\mathbb{W}^\mathsf{T} \cdot \diag{ \mathbf{p} } \cdot ( \mathbb{H}_\textnormal{B}\mathbb{W}^\mathsf{T} )^\mathsf{T} \right|^{\frac{1}{n_\textnormal{B}}} \right), \qquad \\
f_\textnormal{E}(\mathbb{W}) 
&=& \frac{1}{2} \log \left| \mathbb{H}_\textnormal{E}\mathbb{W}^\mathsf{T} \cdot \diag{ \mathbf{v} } \cdot ( \mathbb{H}_\textnormal{E}\mathbb{W}^\mathsf{T} )^\mathsf{T}  + \mathbb{I}_{n_\textnormal{E}}\right|,
\end{IEEEeqnarray}
\end{small}where $f_\textnormal{B}(\mathbb{W}) $ corresponds to Bob and $f_\textnormal{E}(\mathbb{W}) $ to Eve. Then we separately construct the surrogate functions $\widetilde{f}_\textnormal{B}(\mathbb{W};\mathbb{W}^{(k)})$ and $\widetilde{f}_\textnormal{E}(\mathbb{W};\mathbb{W}^{(k)})$ for $f_\textnormal{B}(\mathbb{W})$ and $f_\textnormal{E}(\mathbb{W})$ such that
\begin{IEEEeqnarray}{rCl}
	\widetilde{f}(\mathbb{W};\mathbb{W}^{(k)}) = \widetilde{f}_\textnormal{B}(\mathbb{W};\mathbb{W}^{(k)}) + \widetilde{f}_\textnormal{E}(\mathbb{W};\mathbb{W}^{(k)}). \label{eq: f approx}
\end{IEEEeqnarray}

For $f_\textnormal{B}(\mathbb{W})$, we first characterize its gradient and Hessian matrix in the following lemma, whose proof is given in Appendix~\ref{sec: appendix A}.
\begin{lemma}\label{lemma 2}
\begin{IEEEeqnarray}{rCl}
\nabla f_\textnormal{B}(\mathbb{W}) 
&=& - \frac{ |\mathbb{H}_\textnormal{B} \mathbb{W}^\mathsf{T} \cdot \diag{\mathbf{p}} \cdot \mathbb{W} \mathbb{H}_\textnormal{B}^\mathsf{T} |^{ \frac{1}{n_\textnormal{B}} } }{ 1 + |\mathbb{H}_\textnormal{B} \mathbb{W}^\mathsf{T} \cdot \diag{\mathbf{p}} \cdot \mathbb{W} \mathbb{H}_\textnormal{B}^\mathsf{T} |^{ \frac{1}{n_\textnormal{B}} } } \nonumber\\
&& \quad \cdot \diag{\mathbf{p}} \cdot \mathbb{W} \mathbb{H}_\textnormal{B}^\mathsf{T} ( \mathbb{H}_\textnormal{B} \mathbb{W}^\mathsf{T} \cdot \diag{\mathbf{p}} \cdot \mathbb{W} \mathbb{H}_\textnormal{B}^\mathsf{T} )^{-1}  \nonumber\\
&& \quad \cdot \mathbb{H}_\textnormal{B}, \label{eq: f1 gradient}\\
\mathbb{G}_{f_\textnormal{B}}(\mathbb{W})
&=& \mathbb{U}_1 \otimes \mathbb{V}_1 + \mathbb{K}_{ n_\textnormal{T}\times n_\textnormal{T} } \cdot \mathbb{O}_1^\mathsf{T} \otimes \mathbb{O}_1 - \mathbb{U}_2 \otimes \mathbb{V}_2, \label{eq: f1 hessian} \quad
\end{IEEEeqnarray}
where $\mathbb{K}_{ n_\textnormal{T}\times n_\textnormal{T} }$ is the commutation matrix, and
\begin{IEEEeqnarray}{rCl}
\mathbb{U}_1 & = & \mathbb{U}_2 =\mathbb{H}_\textnormal{B}^\mathsf{T} (\mathbb{H}_\textnormal{B} \mathbb{W}^\mathsf{T} \cdot \diag{\mathbf{p}} \cdot \mathbb{W} \mathbb{H}_\textnormal{B}^\mathsf{T})^{-1} \mathbb{H}_\textnormal{B}, \nonumber \\
\mathbb{V}_1 & = & \diag{\mathbf{p}} \cdot \mathbb{W}\mathbb{H}_\textnormal{B}^\mathsf{T} (\mathbb{H}_\textnormal{B} \mathbb{W}^\mathsf{T} \cdot \diag{\mathbf{p}} \cdot \mathbb{W} \mathbb{H}_\textnormal{B}^\mathsf{T})^{-1} \mathbb{H}_\textnormal{B} \mathbb{W}^\mathsf{T} \nonumber\\
&& \cdot \diag{\mathbf{p}}, \nonumber\\
\mathbb{O}_1 & = & \mathbb{H}_\textnormal{B}^\mathsf{T} (\mathbb{H}_\textnormal{B} \mathbb{W}^\mathsf{T} \cdot \diag{\mathbf{p}} \cdot \mathbb{W} \mathbb{H}_\textnormal{B}^\mathsf{T})^{-1} \mathbb{H}_\textnormal{B} \mathbb{W}^\mathsf{T} \cdot \diag{\mathbf{p}}, \nonumber \\
\mathbb{V}_2 &=& \diag{\mathbf{p}}. \nonumber \hfill\blacksquare
\end{IEEEeqnarray}
\end{lemma}
For convenience, we denote the Hessian matrix of $f_\textnormal{B}(\mathbb{W})$ at $\mathbb{W}^{(k)}$ by $\mathbb{G}_{f_\textnormal{B}}^{(k)}$. With Lemma~\ref{lemma 2}, we construct $\widetilde{f}_\textnormal{B}(\mathbb{W};\mathbb{W}^{(k)})$ based on the second-order Taylor expansion with a positive semi-definite Hessian matrix of $f_\textnormal{B}(\mathbb{W})$ such that
\begin{IEEEeqnarray}{rCl}
&&\widetilde{f}_\textnormal{B}(\mathbb{W};\mathbb{W}^{(k)}) 
= f_\textnormal{B}(\mathbb{W}^{(k)}) + \mathrm{tr}\{ \nabla f_\textnormal{B}(\mathbb{W}^{(k)})^\mathsf{T} ( \mathbb{W}- \mathbb{W}^{(k)}) \} \nonumber\\
&& \qquad\qquad + \frac{1}{2} \mathrm{vec}( \mathbb{W}- \mathbb{W}^{(k)} )^\mathsf{T} \hat{\mathbb{G}}_{f_\textnormal{B}}^{(k)} \mathrm{vec}( \mathbb{W}- \mathbb{W}^{(k)}), \label{eq: f1 approx 1}\qquad
\end{IEEEeqnarray}
where $\hat{\mathbb{G}}_{f_\textnormal{B}}^{(k)}$ is the nearest symmetric positive semi-definite matrix to $\mathbb{G}_{f_\textnormal{B}}^{(k)}$, which achieves the minimum Frobenius-norm error, i.e., $\hat{\mathbb{G}}_{f_\textnormal{B}}^{(k)} = \argmin_{\mathbb{G}_{f_\textnormal{B}}\in \mathcal{S}_{+}^{n}} \|\mathbb{G}_{f_\textnormal{B}} - \mathbb{G}_{f_\textnormal{B}}^{(k)} \|_\mathsf{F}$ with $n=n_\textnormal{T}^2$. It has been shown in \cite{Higham1998} that if the eigenvalue decomposition of $\mathbb{G}_{f_\textnormal{B}}^{(k)}$ is $\mathbb{G}_{f_\textnormal{B}}^{(k)}=\mathbb{P} \cdot \diag{\mathbf{h}} \cdot \mathbb{P}^\mathsf{T}$, then $\hat{\mathbb{G}}_{f_\textnormal{B}}^{(k)}$ can be formulated by $\hat{\mathbb{G}}_{f_\textnormal{B}}^{(k)}=\mathbb{P} \cdot \diag{\mathbf{h}_+} \cdot \mathbb{P}^\mathsf{T}$,
where $\mathbf{h}=(h_1,\cdots,h_{n_\textnormal{T}^2})^\mathsf{T}$, and $\mathbf{h}_+=(h_1^+,\cdots,h_{n_\textnormal{T}^2}^+)^\mathsf{T}$ with $h_i^+=\Bigl\{\begin{array}{l}
h_i, \textnormal{ if } h_i\geq 0,\\
0, \textnormal{  if } h_i< 0.
\end{array}$. To ensure that $\widetilde{f}_\textnormal{B}(\mathbb{W};\mathbb{W}^{(k)}) $ is strongly convex with respect to $\mathbb{W}$, we further include a second-order term to \eqref{eq: f1 approx 1} such that
\begin{IEEEeqnarray}{rCl}
\widetilde{f}_\textnormal{B}(\mathbb{W};\mathbb{W}^{(k)}) 
&=& f_\textnormal{B}(\mathbb{W}^{(k)}) + \mathrm{tr}\{ \nabla f_\textnormal{B}(\mathbb{W}^{(k)})^\mathsf{T} ( \mathbb{W}- \mathbb{W}^{(k)}) \} \nonumber\\
&& + \frac{1}{2} \mathrm{vec}( \mathbb{W}- \mathbb{W}^{(k)} )^\mathsf{T} \hat{\mathbb{G}}_{f_\textnormal{B}}^{(k)} \mathrm{vec}( \mathbb{W}- \mathbb{W}^{(k)}) \nonumber\\
&&+ \frac{\tau}{4} \| \mathbb{W}- \mathbb{W}^{(k)} \|_\mathsf{F}^2, \label{eq: f1 approx}
\end{IEEEeqnarray}
where $\tau>0$.

For $f_\textnormal{E}(\mathbb{W})$, we construct its surrogate function by combining the first-order Taylor expansion of the log-det function:
\begin{IEEEeqnarray}{rCl}
\log|\mathbb{X}| 
&=& \log|\mathbb{X}^{(k)}| + \mathrm{tr}\bigl\{ (\mathbb{X}^{(k)})^{-1} \cdot (\mathbb{X}-\mathbb{X}^{(k)})\bigr\}. \label{eq: log|X| derivate}
\end{IEEEeqnarray}
Substituting $\mathbb{X} =\mathbb{H}_\textnormal{E}\mathbb{W}^\mathsf{T} \cdot \diag{ \mathbf{v} } \cdot ( \mathbb{H}_\textnormal{E}\mathbb{W}^\mathsf{T} )^\mathsf{T}  + \mathbb{I}_{n_\textnormal{E}}$ and $\mathbb{X}^{(k)} =\mathbb{H}_\textnormal{E}(\mathbb{W}^{(k)})^\mathsf{T} \cdot \diag{ \mathbf{v} } \cdot ( \mathbb{H}_\textnormal{E}(\mathbb{W}^{(k)})^\mathsf{T} )^\mathsf{T}  + \mathbb{I}_{n_\textnormal{E}}$ into \eqref{eq: log|X| derivate} and multiplying $\frac{1}{2}$ on both sides, we let
\begin{IEEEeqnarray}{rCl}
\IEEEeqnarraymulticol{3}{l}{%
	\widetilde{f}_\textnormal{E}(\mathbb{W};\mathbb{W}^{(k)})
}\nonumber\\
&=& \frac{1}{2} \log \Bigl|\mathbb{H}_\textnormal{E} (\mathbb{W}^{(k)} )^\mathsf{T} \cdot \diag{ \mathbf{v} } \cdot ( \mathbb{H}_\textnormal{E}(\mathbb{W}^{(k)})^\mathsf{T} )^\mathsf{T}  + \mathbb{I}_{n_\textnormal{E}} \Bigr| \nonumber\\
&& + \frac{1}{2} \mathrm{tr}\Bigl\{ \Bigl( \mathbb{H}_\textnormal{E}(\mathbb{W}^{(k)})^\mathsf{T} \cdot \diag{ \mathbf{v} } \cdot ( \mathbb{H}_\textnormal{E}(\mathbb{W}^{(k)})^\mathsf{T} )^\mathsf{T}  + \mathbb{I}_{n_\textnormal{E}} \Bigr)^{-1} \nonumber\\
&&\qquad \cdot\Bigl( \mathbb{H}_\textnormal{E}\mathbb{W}^\mathsf{T} \cdot \diag{ \mathbf{v} } \cdot ( \mathbb{H}_\textnormal{E}\mathbb{W}^\mathsf{T} )^\mathsf{T} + \mathbb{I}_{n_\textnormal{E} } \Bigr) \Bigr\} -\frac{n_\textnormal{E}}{2}. \label{eq: f2 approx 1}
\end{IEEEeqnarray}
Note that \eqref{eq: f2 approx 1} is a convex function of $\mathbb{W}$. Similar to \eqref{eq: f1 approx}, we also include a second-order term to ensure that $\widetilde{f}_\textnormal{E}(\mathbb{W};\mathbb{W}^{(k)}) $ is strongly convex with respect to $\mathbb{W}$ such that
\begin{IEEEeqnarray}{rCl}
\IEEEeqnarraymulticol{3}{l}{%
	\widetilde{f}_\textnormal{E}(\mathbb{W};\mathbb{W}^{(k)})
}\nonumber\\
&=& \frac{1}{2} \log \Bigl|\mathbb{H}_\textnormal{E} (\mathbb{W}^{(k)} )^\mathsf{T} \cdot \diag{ \mathbf{v} } \cdot ( \mathbb{H}_\textnormal{E}(\mathbb{W}^{(k)})^\mathsf{T} )^\mathsf{T}  + \mathbb{I}_{n_\textnormal{E}} \Bigr| \nonumber\\
&& + \frac{1}{2} \mathrm{tr}\Bigl\{ \Bigl( \mathbb{H}_\textnormal{E}(\mathbb{W}^{(k)})^\mathsf{T} \cdot \diag{ \mathbf{v} } \cdot ( \mathbb{H}_\textnormal{E}(\mathbb{W}^{(k)})^\mathsf{T} )^\mathsf{T}  + \mathbb{I}_{n_\textnormal{E}} \Bigr)^{-1} \nonumber\\
&&\qquad\quad \cdot\Bigl( \mathbb{H}_\textnormal{E}\mathbb{W}^\mathsf{T} \cdot \diag{ \mathbf{v} } \cdot ( \mathbb{H}_\textnormal{E}\mathbb{W}^\mathsf{T} )^\mathsf{T} + \mathbb{I}_{n_\textnormal{E} } \Bigr) \Bigr\} \nonumber\\
&& + \frac{\tau}{4} \| \mathbb{W}- \mathbb{W}^{(k)} \|_\mathsf{F}^2 -\frac{n_\textnormal{E}}{2}. \label{eq: f2 approx}
\end{IEEEeqnarray}

Substituting \eqref{eq: f1 approx} and \eqref{eq: f2 approx} into \eqref{eq: f approx}, we can show that $\widetilde{f}(\mathbb{W}^{(k)};\mathbb{W}^{(k)}) =  f(\mathbb{W}^{(k)})$, $\nabla \widetilde{f}(\mathbb{W}^{(k)};\mathbb{W}^{(k)})= \nabla f(\mathbb{W}^{(k)})$, and the Hessian matrix of $\widetilde{f}(\mathbb{W};\mathbb{W}^{(k)})$ satisfies $\mathbb{G}_{\widetilde{f}}(\mathbb{W};\mathbb{W}^{(k)}) \succcurlyeq \tau\mathbb{I}_{n_\textnormal{T}^2}$, indicating that $\widetilde{f}(\mathbb{W};\mathbb{W}^{(k)})$ is a strongly convex function. Therefore, the fully-connected beamforming optimization problem can be iteratively solved by the following convex subproblem:
\begin{IEEEeqnarray}{rCl}
	\mathcal{P}_{1\& 2}^\prime:\
	&& \hat{\mathbb{W}}^{(k+1)} =\argmin_{\mathbb{W}\in\mathcal{W}_i} \widetilde{f}(\mathbb{W};\mathbb{W}^{(k)}), \label{eq: convex problem}
\end{IEEEeqnarray}
where $\mathcal{W}_i$ could be $\mathcal{W}_1$ or $\mathcal{W}_2$, which corresponds to problem $\mathcal{P}_1$ or $\mathcal{P}_2$, respectively. Besides, the next iteration variable $\mathbb{W}^{(k+1)}$ is generated similar to \eqref{eq: xk update} such that
\begin{IEEEeqnarray}{rCl}
\mathbb{W}^{(k+1)} = \mathbb{W}^{(k)} + \gamma^{(k)} (\hat{\mathbb{W}}^{(k+1)}-\mathbb{W}^{(k)}).\label{eq: Wk update} 
\end{IEEEeqnarray}
We summarize the above proposed SCA-aided algorithm to solve $\mathcal{P}_1$ or $\mathcal{P}_2$ in the following.
\begin{algorithm}[H]
	\caption{}\label{alg: alg1}
	\begin{algorithmic}[1]	
		\STATE {Initialize} $\mathbb{W}^{(0)}$, select a sequence $\{\gamma^{(k)}\}$;
		\STATE \textbf{for} $k=0,1,2,\cdots$ \textbf{do}
		\STATE \hspace{0.25cm} Calculate $\nabla f_\textnormal{B}(\mathbb{W}^{(k)}) $ in \eqref{eq: f1 gradient}, $\mathbb{G}^{(k)}_{f_\textnormal{B}}$ in \eqref{eq: f1 hessian};
		\STATE \hspace{0.25cm} Solve the problem \eqref{eq: convex problem} by CVX to obtain $\hat{\mathbb{W}}^{(k+1)}$;
		\STATE \hspace{0.25cm} Update $\mathbb{W}^{(k+1)}$ by \eqref{eq: Wk update};
		\STATE \hspace{0.25cm} Terminate loop if converges.
	\end{algorithmic}
\end{algorithm}

\subsubsection{Sub-connected Beamforming Optimization Algorithm}
The procedure of solving problems $\mathcal{P}_3$ or $\mathcal{P}_4$ is similar to Sec.~\ref{full-connect algorithm}. Here we mainly emphasize the differences. Given a fixed $\mathcal{I}$ satisfying $\mathcal{I}\in\mathcal{I}_\textnormal{all}$, we also denote the objective function of problem $\mathcal{P}_3$ or $\mathcal{P}_4$ by $-f(\mathbb{B})$, i.e.,
\begin{align}
 f(\mathbb{B}) = - \mathsf{R}_\textnormal{s}^\textnormal{I}(\mathbb{H}_{ \textnormal{B}, \mathcal{I} }+\mathbb{H}_{ \textnormal{B}, \mathcal{I}^\textnormal{c} } \mathbb{B}^\mathsf{T},\mathbb{H}_{ \textnormal{E}, \mathcal{I} }+\mathbb{H}_{ \textnormal{E}, \mathcal{I}^\textnormal{c} } \mathbb{B}^\mathsf{T} ).
\end{align}
Denote the feasible domain of problem $\mathcal{P}_3$ by $\mathcal{B}_1$, and the feasible domain of problem $\mathcal{P}_4$ by $\mathcal{B}_2$. We divide $f(\mathbb{B})$ into two parts such that $f(\mathbb{B})=f_\textnormal{B}(\mathbb{B})+f_\textnormal{E}(\mathbb{B})$ with 
\begin{IEEEeqnarray}{rCl}
f_\textnormal{B}(\mathbb{B}) 
&=& - \frac{n_\textnormal{B}}{2} \log \Bigl( 1 + \bigl| ( \mathbb{H}_{\textnormal{B},\mathcal{I}} +\mathbb{H}_{\textnormal{B},\mathcal{I}^\textnormal{c} } \mathbb{B}^\mathsf{T} ) \cdot \diag{ \mathbf{p}_\mathcal{I} } \nonumber\\
&& \qquad\qquad\qquad \cdot ( \mathbb{H}_{\textnormal{B},\mathcal{I}} +\mathbb{H}_{\textnormal{B},\mathcal{I}^\textnormal{c} } \mathbb{B}^\mathsf{T} )^\mathsf{T} \bigr|^{\frac{1}{n_\textnormal{B}}} \Bigr), \qquad \\
f_\textnormal{E}(\mathbb{B}) 
&=& \frac{1}{2} \log \Bigl| ( \mathbb{H}_{\textnormal{E},\mathcal{I}} +\mathbb{H}_{\textnormal{E},\mathcal{I}^\textnormal{c} } \mathbb{B}^\mathsf{T} ) \cdot \diag{ \mathbf{v}_\mathcal{I} } \nonumber\\
&& \qquad\ \cdot ( \mathbb{H}_{\textnormal{E},\mathcal{I}} +\mathbb{H}_{\textnormal{E},\mathcal{I}^\textnormal{c} } \mathbb{B}^\mathsf{T} )^\mathsf{T}  + \mathbb{I}_{n_\textnormal{E}} \Bigr|, 
\end{IEEEeqnarray}
where $\mathbf{p}_\mathcal{I}$ and $\mathbf{v}_\mathcal{I}$ denote the subvectors of $\mathbf{p}$ and $\mathbf{v}$ with entries indexed by $\mathcal{I}$, respectively.

For $f_\textnormal{B}(\mathbb{B}) $, its gradient and Hessian matrix are given in the following lemma whose proof is similar to Appendix~\ref{sec: appendix A} and omitted.
\begin{lemma}\label{lemma 3}
\begin{small}	
\begin{IEEEeqnarray}{rCl}
	\IEEEeqnarraymulticol{3}{l}{%
		\nabla f_\textnormal{B}(\mathbb{B})
	}\nonumber\\
	 &=& {\scriptsize{ - \frac{ | ( \mathbb{H}_{\textnormal{B},\mathcal{I}} +\mathbb{H}_{\textnormal{B},\mathcal{I}^\textnormal{c} } \mathbb{B}^\mathsf{T} ) \cdot \diag{ \mathbf{p}_\mathcal{I} } \cdot ( \mathbb{H}_{\textnormal{B},\mathcal{I}} +\mathbb{H}_{\textnormal{B},\mathcal{I}^\textnormal{c} } \mathbb{B}^\mathsf{T} )^\mathsf{T} |^{ \frac{1}{n_\textnormal{B}} } }{ 1 + | ( \mathbb{H}_{\textnormal{B},\mathcal{I}} +\mathbb{H}_{\textnormal{B},\mathcal{I}^\textnormal{c} } \mathbb{B}^\mathsf{T} ) \cdot \diag{\mathbf{p}_\mathcal{I} } \cdot ( \mathbb{H}_{\textnormal{B},\mathcal{I}} +\mathbb{H}_{\textnormal{B},\mathcal{I}^\textnormal{c} } \mathbb{B}^\mathsf{T} )^\mathsf{T} |^{ \frac{1}{n_\textnormal{B}} } } } }\nonumber\\
	 && \quad \cdot \diag{\mathbf{p}_\mathcal{I} } \cdot ( \mathbb{H}_{\textnormal{B},\mathcal{I}} +\mathbb{H}_{\textnormal{B},\mathcal{I}^\textnormal{c} } \mathbb{B}^\mathsf{T} )^\mathsf{T} \Bigl( ( \mathbb{H}_{\textnormal{B},\mathcal{I}} +\mathbb{H}_{\textnormal{B},\mathcal{I}^\textnormal{c} } \mathbb{B}^\mathsf{T} ) \nonumber\\
	&& \quad \cdot \diag{\mathbf{p}_\mathcal{I}} \cdot ( \mathbb{H}_{\textnormal{B},\mathcal{I}} +\mathbb{H}_{\textnormal{B},\mathcal{I}^\textnormal{c} } \mathbb{B}^\mathsf{T} )^\mathsf{T} \Bigr)^{-1} \mathbb{H}_{\textnormal{B},\mathcal{I}^\textnormal{c} }, \label{eq: g1 gradient}\\
	\IEEEeqnarraymulticol{3}{l}{%
		\mathbb{G}_{f_\textnormal{B}}(\mathbb{B})= 
			\mathbb{U}_3 \otimes \mathbb{V}_3  
			+ \mathbb{K}_{ (n_\textnormal{T} -n_\textnormal{B})\times n_\textnormal{B} } \cdot \mathbb{O}_2^\mathsf{T} \otimes \mathbb{O}_2 
			- \mathbb{U}_4 \otimes \mathbb{V}_4 , 	\label{eq: H_g1}
	}
\end{IEEEeqnarray}
\end{small}where 
\begin{IEEEeqnarray}{rCl}
\mathbb{U}_3 & = & \mathbb{U}_4 = \mathbb{H}_{\textnormal{B},\mathcal{I}^\textnormal{c}}^\mathsf{T} \Bigl( ( \mathbb{H}_{\textnormal{B},\mathcal{I}} +\mathbb{H}_{\textnormal{B},\mathcal{I}^\textnormal{c} } \mathbb{B}^\mathsf{T} ) \cdot \diag{\mathbf{p}_\mathcal{I}}  \nonumber\\
&& \cdot ( \mathbb{H}_{\textnormal{B},\mathcal{I}} +\mathbb{H}_{\textnormal{B},\mathcal{I}^\textnormal{c} } \mathbb{B}^\mathsf{T} )^\mathsf{T} \Bigr)^{-1} \mathbb{H}_{\textnormal{B},\mathcal{I}^\textnormal{c}}, \nonumber \\
\mathbb{V}_3 & = & \diag{\mathbf{p}_\mathcal{I}} \cdot ( \mathbb{H}_{\textnormal{B},\mathcal{I}} +\mathbb{H}_{\textnormal{B},\mathcal{I}^\textnormal{c} } \mathbb{B}^\mathsf{T} )^\mathsf{T} \Bigl( ( \mathbb{H}_{\textnormal{B},\mathcal{I}} +\mathbb{H}_{\textnormal{B},\mathcal{I}^\textnormal{c} } \mathbb{B}^\mathsf{T} ) \nonumber\\
&& \cdot \diag{\mathbf{p}_\mathcal{I}} \cdot ( \mathbb{H}_{\textnormal{B},\mathcal{I}} +\mathbb{H}_{\textnormal{B},\mathcal{I}^\textnormal{c} } \mathbb{B}^\mathsf{T} )^\mathsf{T} \Bigr)^{-1} ( \mathbb{H}_{\textnormal{B},\mathcal{I}} +\mathbb{H}_{\textnormal{B},\mathcal{I}^\textnormal{c} } \mathbb{B}^\mathsf{T} ) \nonumber\\
&& \cdot \diag{\mathbf{p}_\mathcal{I}}, \nonumber\\
\mathbb{O}_2 & = & \mathbb{H}_{\textnormal{B},\mathcal{I}^\textnormal{c}}^\mathsf{T} \Bigl( ( \mathbb{H}_{\textnormal{B},\mathcal{I}} +\mathbb{H}_{\textnormal{B},\mathcal{I}^\textnormal{c} } \mathbb{B}^\mathsf{T} ) \cdot \diag{\mathbf{p}_\mathcal{I}} \nonumber\\
&& \cdot ( \mathbb{H}_{\textnormal{B},\mathcal{I}} +\mathbb{H}_{\textnormal{B},\mathcal{I}^\textnormal{c} } \mathbb{B}^\mathsf{T} )^\mathsf{T} \Bigr)^{-1} ( \mathbb{H}_{\textnormal{B},\mathcal{I}} +\mathbb{H}_{\textnormal{B},\mathcal{I}^\textnormal{c} } \mathbb{B}^\mathsf{T} ) \cdot \diag{\mathbf{p}_\mathcal{I}}, \nonumber \\
\mathbb{V}_4 &=& \diag{\mathbf{p}_\mathcal{I}}. \nonumber \hfill\blacksquare
\end{IEEEeqnarray}
\end{lemma}
Combined with Lemma~\ref{lemma 3}, we let
\begin{IEEEeqnarray}{rCl}
\widetilde{f}_\textnormal{B}(\mathbb{B};\mathbb{B}^{(k)}) 
&=& f_\textnormal{B}(\mathbb{B}^{(k)}) + \mathrm{tr}\{ \nabla f_\textnormal{B}(\mathbb{B}^{(k)})^\mathsf{T} ( \mathbb{B}- \mathbb{B}^{(k)}) \} \nonumber\\
&& + \frac{1}{2} \mathrm{vec}( \mathbb{B}- \mathbb{B}^{(k)} )^\mathsf{T} \hat{\mathbb{G}}_{f_\textnormal{B}}^{(k)} \mathrm{vec}( \mathbb{B}- \mathbb{B}^{(k)}) \nonumber\\
&+& \frac{\tau}{4} \| \mathbb{B}- \mathbb{B}^{(k)} \|_\mathsf{F}^2, \label{eq: g1 approx}
\end{IEEEeqnarray}
where $\mathbb{G}_{f_\textnormal{B}}^{(k)}$ is the Hessian matrix of $f_\textnormal{B}(\mathbb{B})$ at $\mathbb{B}^{(k)}$, and $\hat{\mathbb{G}}_{f_\textnormal{B}}^{(k)}$ is the nearest symmetric positive semi-definite matrix to $\mathbb{G}_{f_\textnormal{B}}^{(k)}$ achieving the minimum Frobenius-norm error. 

For $f_\textnormal{E}(\mathbb{B})$, we substitute $\mathbb{X} =( \mathbb{H}_{ \textnormal{E}, \mathcal{I} }+\mathbb{H}_{ \textnormal{E}, \mathcal{I}^\textnormal{c} } \mathbb{B}^\mathsf{T} ) \cdot \diag{ \mathbf{v}_\mathcal{I} } \cdot ( \mathbb{H}_{ \textnormal{E}, \mathcal{I} }+\mathbb{H}_{ \textnormal{E}, \mathcal{I}^\textnormal{c} } \mathbb{B}^\mathsf{T} )^\mathsf{T}  + \mathbb{I}_{n_\textnormal{E}}$ and $\mathbb{X}^{(k)} =( \mathbb{H}_{ \textnormal{E}, \mathcal{I} }+\mathbb{H}_{ \textnormal{E}, \mathcal{I}^\textnormal{c} } (\mathbb{B}^{(k)})^\mathsf{T} ) \cdot \diag{ \mathbf{v}_\mathcal{I} } \cdot ( \mathbb{H}_{ \textnormal{E}, \mathcal{I} }+\mathbb{H}_{ \textnormal{E}, \mathcal{I}^\textnormal{c} } (\mathbb{B}^{(k)})^\mathsf{T} )^\mathsf{T}  + \mathbb{I}_{n_\textnormal{E}}$ into \eqref{eq: log|X| derivate}, and let
\begin{IEEEeqnarray}{rCl}
\widetilde{f}_\textnormal{E}(\mathbb{B};\mathbb{B}^{(k)})
&=& \frac{1}{2} \log \Bigl| ( \mathbb{H}_{ \textnormal{E}, \mathcal{I} }+\mathbb{H}_{ \textnormal{E}, \mathcal{I}^\textnormal{c} } (\mathbb{B}^{(k)})^\mathsf{T} ) \cdot \diag{ \mathbf{v}_\mathcal{I} } \nonumber\\
&& \qquad\ \cdot ( \mathbb{H}_{ \textnormal{E}, \mathcal{I} }+\mathbb{H}_{ \textnormal{E}, \mathcal{I}^\textnormal{c} } (\mathbb{B}^{(k)})^\mathsf{T} )^\mathsf{T}  + \mathbb{I}_{n_\textnormal{E}} \Bigr| \nonumber\\
&& + \frac{1}{2} \mathrm{tr}\Bigl\{ \Bigl( ( \mathbb{H}_{ \textnormal{E}, \mathcal{I} }+\mathbb{H}_{ \textnormal{E}, \mathcal{I}^\textnormal{c} } (\mathbb{B}^{(k)})^\mathsf{T} ) \cdot \diag{ \mathbf{v}_\mathcal{I} } \qquad \nonumber\\
&& \qquad\quad \cdot ( \mathbb{H}_{ \textnormal{E}, \mathcal{I} }+\mathbb{H}_{ \textnormal{E}, \mathcal{I}^\textnormal{c} } (\mathbb{B}^{(k)})^\mathsf{T} )^\mathsf{T}  + \mathbb{I}_{n_\textnormal{E}} \Bigr)^{-1} \nonumber\\
&&\qquad\quad \cdot \Bigl( ( \mathbb{H}_{ \textnormal{E}, \mathcal{I} }+\mathbb{H}_{ \textnormal{E}, \mathcal{I}^\textnormal{c} } \mathbb{B}^\mathsf{T} ) \cdot \diag{ \mathbf{v}_\mathcal{I} } \nonumber\\
&&\qquad\quad \cdot ( \mathbb{H}_{ \textnormal{E}, \mathcal{I} }+\mathbb{H}_{ \textnormal{E}, \mathcal{I}^\textnormal{c} } \mathbb{B}^\mathsf{T} )^\mathsf{T} +\mathbb{I}_{n_\textnormal{E}} \Bigr) \Bigr\} \nonumber\\
&& + \frac{\tau}{4} \| \mathbb{B}- \mathbb{B}^{(k)} \|_\mathsf{F}^2 -\frac{n_\textnormal{E}}{2}. \label{eq: g2 approx}
\end{IEEEeqnarray}

As a result, we construct the surrogate function $\widetilde{f}(\mathbb{B};\mathbb{B}^{(k)})$ for $f(\mathbb{B})$ at $k$-th iteration by
\begin{IEEEeqnarray}{rCl}
\widetilde{f}(\mathbb{B};\mathbb{B}^{(k)}) = \widetilde{f}_\textnormal{B}(\mathbb{B};\mathbb{B}^{(k)}) + \widetilde{f}_\textnormal{E}(\mathbb{B};\mathbb{B}^{(k)}).
\end{IEEEeqnarray}
Combining \eqref{eq: g1 approx} and \eqref{eq: g2 approx}, we can show that the Hessian matrix of $\widetilde{f}(\mathbb{B};\mathbb{B}^{(k)})$ satisfies $\mathbb{G}_{\widetilde{f}}(\mathbb{B};\mathbb{B}^{(k)}) \succcurlyeq \tau\mathbb{I}_{n_\textnormal{B}\cdot(n_\textnormal{T}-n_\textnormal{B})}, \forall \mathbb{B}\in\mathcal{B}_i$. Therefore, the strongly convex subproblem at $k$ iteration is given by
\begin{IEEEeqnarray}{rCl}
	\mathcal{P}_{3\& 4}^\prime:\
	&& \hat{\mathbb{B}}^{(k+1)} =\argmin_{\mathbb{B}\in\mathcal{B}_i} \widetilde{f}(\mathbb{B};\mathbb{B}^{(k)}), \label{eq: convex problem for g}
\end{IEEEeqnarray}
where $\mathcal{B}_i$ could be $\mathcal{B}_1$ or $\mathcal{B}_2$. And the next iteration variable $\mathbb{B}^{(k+1)}$ is generated by
\begin{IEEEeqnarray}{rCl}
\mathbb{B}^{(k+1)} = \mathbb{B}^{(k)} + \gamma^{(k)} (\hat{\mathbb{B}}^{(k+1)}-\mathbb{B}^{(k)}). \label{eq: Bk update}
\end{IEEEeqnarray}
We summarize the above proposed SCA-aided algorithm for problem $\mathcal{P}_3$ or $\mathcal{P}_4$ in the following.
\begin{algorithm}[H]
	\caption{}\label{alg: alg2}
	\begin{algorithmic}[1]	
		\STATE \textbf{for} $\mathcal{I}\in\mathcal{I}_\textnormal{all}$ \textbf{do}
		\STATE \hspace{0.2cm} {Initialize} $\mathbb{B}^{(0)}$, select a sequence $\{\gamma^{(k)}\}$;
		\STATE \hspace{0.2cm} \textbf{for} $k=0,1,2,\cdots$ \textbf{do}
		\STATE \hspace{0.5cm} Calculate $\nabla f_\textnormal{B}(\mathbb{B}^{(k)}) $ in \eqref{eq: g1 gradient}, $\mathbb{G}^{(k)}_{f_\textnormal{B}}$ in \eqref{eq: H_g1};
		\STATE \hspace{0.5cm} Solve the problem \eqref{eq: convex problem for g} by CVX to obtain $\hat{\mathbb{B}}^{(k+1)}$;
		\STATE \hspace{0.5cm} Update $\mathbb{B}^{(k+1)}$ by \eqref{eq: Bk update};
		\STATE \hspace{0.5cm} Terminate loop if converges;
		\STATE \hspace{0.2cm} \textbf{end}
		\STATE \hspace{0.2cm} Save $\mathbb{B}^{(k+1)}$ as $\mathbb{B}_{\mathcal{I}}$; 
		\STATE \textbf{end}
		\STATE Output the optimal beamformer $\mathbb{B}^\star$ among $\mathbb{B}_\mathcal{I}$s, $\mathcal{I}\in\mathcal{I}_\textnormal{all}$.
	\end{algorithmic}
\end{algorithm}

\subsection{Algorithm for Case II} 
For problem $\mathcal{P}_6$, we also denote its objective function by $-f(\mathbb{W})$ and feasible domain by $\mathcal{W}_3$. Similarly, We expand $f(\mathbb{W})=f_\textnormal{B}(\mathbb{W})+f_\textnormal{E}(\mathbb{W})$ with 
\begin{IEEEeqnarray}{rCl}
f_\textnormal{B}(\mathbb{W}) 
&=& - \frac{n_\textnormal{T}}{2} \log \left( 1 + |\diag{ \mathbf{p} }|^{\frac{1}{n_\textnormal{T}}}  \cdot \left| \mathbb{W} \mathbb{H}_\textnormal{B}^\mathsf{T} \mathbb{H}_\textnormal{B}\mathbb{W}^\mathsf{T} \right|^{\frac{1}{n_\textnormal{T}}} \right), \qquad \\
f_\textnormal{E}(\mathbb{W}) 
&=& \frac{1}{2} \log \left| \diag{ \mathbf{v} } \cdot \mathbb{W} \mathbb{H}_\textnormal{E}^\mathsf{T} \mathbb{H}_\textnormal{E}\mathbb{W}^\mathsf{T} + \mathbb{I}_{n_\textnormal{T}} \right|.
\end{IEEEeqnarray}

For $f_\textnormal{B}(\mathbb{W}) $, its gradient and Hessian matrix are given in the following lemma whose proof is similar to Appendix~\ref{sec: appendix A} and omitted.
\begin{lemma}\label{lemma 5}
\begin{IEEEeqnarray}{rCl}
	\nabla f_\textnormal{B}(\mathbb{W})
	 &=& - \frac{ |\diag{\mathbf{p}}|^{\frac{1}{n_\textnormal{T}}} | \mathbb{W} \mathbb{H}_\textnormal{B}^\mathsf{T} \mathbb{H}_\textnormal{B} \mathbb{W}^\mathsf{T} |^{ \frac{1}{n_\textnormal{T}} } }{ 1 + |\diag{\mathbf{p}}|^{\frac{1}{n_\textnormal{T}}} | \mathbb{W} \mathbb{H}_\textnormal{B}^\mathsf{T} \mathbb{H}_\textnormal{B} \mathbb{W}^\mathsf{T} |^{ \frac{1}{n_\textnormal{T}} } }\nonumber\\
	 && \quad \cdot ( \mathbb{W} \mathbb{H}_\textnormal{B}^\mathsf{T} \mathbb{H}_\textnormal{B} \mathbb{W}^\mathsf{T} )^{-1} \mathbb{W} \mathbb{H}_\textnormal{B}^\mathsf{T} \mathbb{H}_\textnormal{B}, \label{eq: s1 gradient}\\
	\mathbb{G}_{f_\textnormal{B}}(\mathbb{W})
	&=& -
	\mathbb{U}_5 \otimes \mathbb{V}_5  
	+ \mathbb{K}_{ n_\textnormal{T} \times n_\textnormal{T} } \cdot \mathbb{O}_3^\mathsf{T} \otimes \mathbb{O}_3 
	+ \mathbb{U}_6 \otimes \mathbb{V}_6 ,\qquad \label{eq: H_s1}
\end{IEEEeqnarray}
where 
\begin{IEEEeqnarray}{rCl}
\mathbb{U}_5 & = & \mathbb{H}_\textnormal{B}^\mathsf{T} \mathbb{H}_\textnormal{B}, \nonumber\\
\mathbb{V}_5 & = & \mathbb{V}_6 = ( \mathbb{W} \mathbb{H}_\textnormal{B}^\mathsf{T} \mathbb{H}_\textnormal{B} \mathbb{W}^\mathsf{T} )^{-1}, \nonumber\\
\mathbb{O}_3 & = & \mathbb{H}_\textnormal{B}^\mathsf{T} \mathbb{H}_\textnormal{B} \mathbb{W}^\mathsf{T} ( \mathbb{W} \mathbb{H}_\textnormal{B}^\mathsf{T} \mathbb{H}_\textnormal{B} \mathbb{W}^\mathsf{T} )^{-1}, \nonumber\\
\mathbb{U}_6 &=& \mathbb{H}_\textnormal{B}^\mathsf{T} \mathbb{H}_\textnormal{B} \mathbb{W}^\mathsf{T} ( \mathbb{W} \mathbb{H}_\textnormal{B}^\mathsf{T} \mathbb{H}_\textnormal{B} \mathbb{W}^\mathsf{T} )^{-1} \mathbb{W} \mathbb{H}_\textnormal{B}^\mathsf{T} \mathbb{H}_\textnormal{B}. \qquad\qquad \qquad\nonumber \hfill\blacksquare
\end{IEEEeqnarray}
\end{lemma}
Combined with Lemma~\ref{lemma 5}, we design
\begin{IEEEeqnarray}{rCl}
\widetilde{f}_\textnormal{B}(\mathbb{W};\mathbb{W}^{(k)}) 
&=& f_\textnormal{B}(\mathbb{W}^{(k)}) + \mathrm{tr}\{ \nabla f_\textnormal{B}(\mathbb{W}^{(k)})^\mathsf{T} ( \mathbb{W}- \mathbb{W}^{(k)}) \} \nonumber\\
&& + \frac{1}{2} \mathrm{vec}( \mathbb{W}- \mathbb{W}^{(k)} )^\mathsf{T} \hat{\mathbb{G}}_{f_\textnormal{B}}^{(k)} \mathrm{vec}( \mathbb{W}- \mathbb{W}^{(k)}) \nonumber\\
&+& \frac{\tau}{4} \| \mathbb{W}- \mathbb{W}^{(k)} \|_\mathsf{F}^2, \label{eq: s1 approx}
\end{IEEEeqnarray}
where $\mathbb{G}_{f_\textnormal{B}}^{(k)}$ is the Hessian matrix of $f_\textnormal{B}(\mathbb{W})$ at $\mathbb{W}^{(k)}$, and $\hat{\mathbb{G}}_{f_\textnormal{B}}^{(k)}$ is the nearest symmetric positive semi-definite matrix to $\mathbb{G}_{f_\textnormal{B}}^{(k)}$ achieving the minimum Frobenius-norm error. 

For $f_\textnormal{E}(\mathbb{W})$, we let
\begin{IEEEeqnarray}{rCl}
\IEEEeqnarraymulticol{3}{l}{%
	\widetilde{f}_\textnormal{E}(\mathbb{W};\mathbb{W}^{(k)})
}\nonumber\\ \quad
&=& \frac{1}{2} \log \Bigl| \diag{ \mathbf{v} } \cdot \mathbb{W}^{(k)} \mathbb{H}_\textnormal{E}^\mathsf{T} \mathbb{H}_\textnormal{E} (\mathbb{W}^{(k)})^\mathsf{T} + \mathbb{I}_{n_\textnormal{T}} \Bigr| \nonumber\\
&& + \frac{1}{2} \mathrm{tr}\Bigl\{ \Bigl( \diag{ \mathbf{v} } \cdot \mathbb{W}^{(k)} \mathbb{H}_\textnormal{E}^\mathsf{T} \mathbb{H}_\textnormal{E} (\mathbb{W}^{(k)})^\mathsf{T} + \mathbb{I}_{n_\textnormal{T}} \Bigr)^{-1} \nonumber\\
&&\qquad\quad \cdot \Bigl( \diag{ \mathbf{v} } \cdot \mathbb{W} \mathbb{H}_\textnormal{E}^\mathsf{T} \mathbb{H}_\textnormal{E}\mathbb{W}^\mathsf{T} +\mathbb{I}_{n_\textnormal{T}} \Bigr) \Bigr\} \nonumber\\
&& + \frac{\tau}{4} \| \mathbb{W}- \mathbb{W}^{(k)} \|_\mathsf{F}^2 -\frac{n_\textnormal{T}}{2}. \label{eq: s2 approx}
\end{IEEEeqnarray}

We construct the surrogate function $\widetilde{f}(\mathbb{W};\mathbb{W}^{(k)})$ for $f(\mathbb{W})$ by
\begin{IEEEeqnarray}{rCl}
\widetilde{f}(\mathbb{W};\mathbb{W}^{(k)}) = \widetilde{f}_\textnormal{B}(\mathbb{W};\mathbb{W}^{(k)}) + \widetilde{f}_\textnormal{E}(\mathbb{W};\mathbb{W}^{(k)}),
\end{IEEEeqnarray}
Then, the strongly convex subproblem at $k$ iteration is
\begin{IEEEeqnarray}{rCl}
	\mathcal{P}_{6}^\prime:\
	&& \hat{\mathbb{W}}^{(k+1)} =\argmin_{\mathbb{W}\in\mathcal{W}_3} \widetilde{f}(\mathbb{W};\mathbb{W}^{(k)}). \label{eq: convex problem for s}
\end{IEEEeqnarray}
And the next iteration variable $\mathbb{W}^{(k+1)}$ is generated similar to \eqref{eq: Wk update}. The proposed SCA-aided algorithm for problem $\mathcal{P}_6$ is summarized as follows.
\begin{algorithm}[H]
	\caption{}\label{alg: alg3}
	\begin{algorithmic}[1]	
		\STATE {Initialize} $\mathbb{W}^{(0)}$, select a sequence $\{\gamma^{(k)}\}$;
		\STATE \textbf{for} $k=0,1,2,\cdots$ \textbf{do}
		\STATE \hspace{0.25cm} Calculate $\nabla f_\textnormal{B}(\mathbb{W}^{(k)}) $ in \eqref{eq: s1 gradient}, $\mathbb{G}^{(k)}_{f_\textnormal{B}}$ in \eqref{eq: H_s1};
		\STATE \hspace{0.25cm} Solve the problem \eqref{eq: convex problem for s} by CVX to obtain $\hat{\mathbb{W}}^{(k+1)}$;
		\STATE \hspace{0.25cm} Update $\mathbb{W}^{(k+1)}$ by \eqref{eq: Wk update};
		\STATE \hspace{0.25cm} Terminate loop if converges.
	\end{algorithmic}
\end{algorithm}

\subsection{Complexity Analysis} 
Since the proposed Algorithms~\ref{alg: alg1}-\ref{alg: alg3} are all iterative methods, we only discuss the computational complexity in constructing the convex subproblem at each iteration (note that the whole computational complexity of solving such a subproblem depends on the specific CVX solver). {For Algorithm~\ref{alg: alg1}, the main computational costs are from computing the gradient $\nabla f_\textnormal{B}(\mathbb{W}^{(k)})$ and Hessian matrix $\mathbb{G}^{(k)}_{f_\textnormal{B}}$, which are $\mathcal{O}(n_\textnormal{B} n_\textnormal{T}^2 )$ and $\mathcal{O}(n_\textnormal{T}^4)$, respectively.} For Algorithm~\ref{alg: alg2}, the computational costs of computing the gradient $\nabla f_\textnormal{B}(\mathbb{B}^{(k)})$ and Hessian matrix $\mathbb{G}^{(k)}_{f_\textnormal{B}}$ are $\mathcal{O}\bigl( n_\textnormal{B}^3+n_\textnormal{B}^2(n_\textnormal{T}- n_\textnormal{B}) \bigr)$ and $\mathcal{O} \bigl( n_\textnormal{B}^2(n_\textnormal{T}- n_\textnormal{B})^2 \bigr)$, respectively. For Algorithm~\ref{alg: alg3}, the computational costs of computing the gradient $\nabla f_\textnormal{B}(\mathbb{W}^{(k)})$ and Hessian matrix $\mathbb{G}^{(k)}_{f_\textnormal{B}}$ are $\mathcal{O}(n_\textnormal{T} n_\textnormal{B}^2)$ and $\mathcal{O}(n_\textnormal{T}^4)$, respectively.

\section{Simulation Results}\label{sec: simulations}
This section presents simulations evaluating the achievable secrecy rates under various system configurations. First, the analytical expressions derived in Thm.~\ref{thm1} are validated by comparing them against prior work considering a single peak-intensity constraint. Subsequently, the achievable secrecy rates under both peak- and average-intensity constraints are examined, and the performance gains achieved by the proposed beamforming schemes are demonstrated. For clarity and reproducibility, {two sets of randomly generated channel matrices are considered in the simulations}, i.e.,
\begin{itemize}
	\item Group $1$:	
	\begin{IEEEeqnarray}{rCl}	
		\mathbb{H}_\textnormal{B}^{1\times4} &=& 
		{\small{ \bigl(\begin{array}{cccc}
		0.8143  &  0.2435  &  0.9293  &  0.3500
		\end{array}\bigr)}}, \\
		\mathbb{H}_\textnormal{E}^{1\times4} &=& 
		{\small{ \bigl(\begin{array}{cccc}
		0.3034  &  0.2489  &  0.1160  &  0.0267
		\end{array}\bigr)}};\qquad
	\end{IEEEeqnarray}
	\item Group $2$:		
		\begin{IEEEeqnarray}{rCl}
		\mathbb{H}_\textnormal{B}^{2\times4}
		&=& {\small{ \left(\begin{array}{cccc}
		0.9218  &  1.2922 &   1.1557   & 1.3491\\
		1.4157  &  1.4595  &  0.5357  &  1.4340
		\end{array}\right)}}, \\
		\mathbb{H}_\textnormal{E}^{2\times4}
		&=& {\small{\left(\begin{array}{cccc}
		0.1787  &  0.2431 &   0.1555 &   0.2060\\
		0.2577 &   0.1078 &   0.3288 &   0.4682
		\end{array}\right)}}.\qquad
		\end{IEEEeqnarray}
\end{itemize}



\subsection{Single Peak-intensity Constraint}
When $\alpha_i=0.5$ for all $i$, the average-intensity constraint becomes inactive \cite{Lapidoth2009,Chaaban2016,Moser2017}, and the considered system reduces to a scenario with a single peak-intensity constraint. Under this setting, Fig.~\ref{fig: Nt2} compares the achievable secrecy rates from Thm.~\ref{thm1} at $\alpha_i=0.5$ with those reported in prior works \cite{Mostafa2015,Arfaoui2020-twc},\footnote{For Fig.~\ref{fig: Nt2}(b), simulations adopt the parameters specified in \cite{Arfaoui2020-twc}, i.e., ``$K_u=\frac{n_\textnormal{TX}}{L}$, $\alpha=\amp$, $\beta=2$''.} both of which consider a single peak-intensity constraint (with \cite{Mostafa2015} focusing on the MISO channel and \cite{Arfaoui2020-twc} on the MIMO channel). 

As shown in Fig.~\ref{fig: Nt2}(a), the secrecy rate in Thm.~\ref{thm1} at $\alpha_i=0.5$ aligns with that of \cite{Mostafa2015}, thereby validating the correctness of the derived analytical expression. Furthermore, a comparative analysis of Fig.~\ref{fig: Nt2}(a) and Fig.~\ref{fig: Nt2}(b) reveals distinct performance trends across the SNR range. At low SNR, the achievable secrecy rate in \cite{Arfaoui2020-twc} exceeds that of both Thm.~\ref{thm1} and \cite{Mostafa2015}. In contrast, at high SNR, Thm.~\ref{thm1} outperforms \cite{Arfaoui2020-twc}, yielding a strictly higher secrecy rate. These performance trends are attributable to the distinct input distributions employed. Specifically, \cite{Mostafa2015} and Thm.~\ref{thm1} leverage a uniform distribution (as the truncated exponential distribution in \eqref{eq: texp pdf} degenerates to a uniform distribution when $\alpha_i=0.5$), whereas \cite{Arfaoui2020-twc} adopts a discrete distribution. As established in \cite{Lapidoth2009}, for peak-intensity-constrained channels, the optimal input distribution is discrete at low SNR but converges to a uniform distribution at high SNR. This theoretical insight fully accounts for the observed behaviors illustrated in Fig. \ref{fig: Nt2}.

\begin{figure}[!htbp]
	\centering
	\includegraphics[width=3.5in]{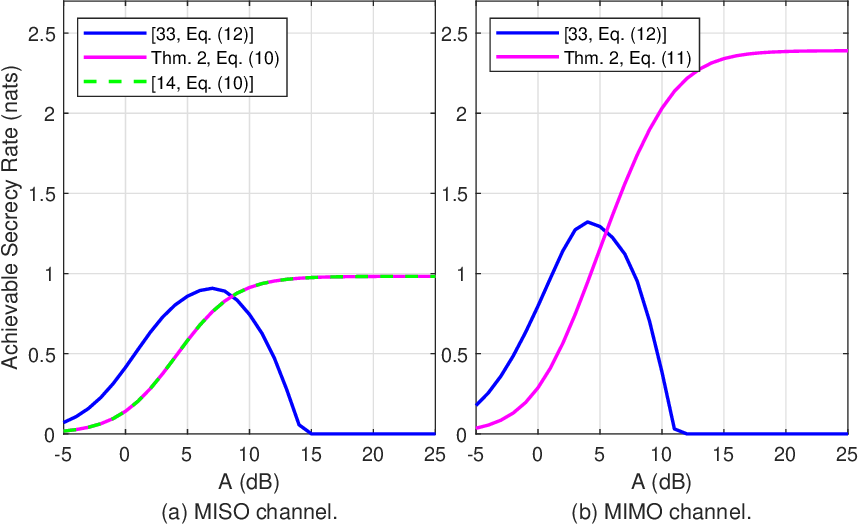}
	\caption{{Comparison of achievable secrecy rates: (a). $\mathbb{H}_\textnormal{B}=\mathbb{H}_\textnormal{B}^{1\times4}$ and $\mathbb{H}_\textnormal{E}=\mathbb{H}_\textnormal{E}^{1\times4}$; (b). $\mathbb{H}_\textnormal{B}=(\mathbb{H}_\textnormal{B}^{2\times4})^\mathsf{T}$ and $\mathbb{H}_\textnormal{E}=(\mathbb{H}_\textnormal{E}^{2\times4})^\mathsf{T}$.}}
	\label{fig: Nt2}
\end{figure}

\vspace{-6mm}
\subsection{Both Peak- and Average-intensity Constraints}
\subsubsection{Fully-connected Beamforming Schemes in $\mathcal{P}_1$ and $\mathcal{P}_2$}\label{sec: full-connect simulation}

We employ Algorithm~\ref{alg: alg1} to solve the fully-connected beamforming optimization problems $\mathcal{P}_1$ and $\mathcal{P}_2$. To do it, we initialize
$\mathbb{W}^{(0)}=\left( \begin{array}{ll}
	\mathbb{I}_{n_\textnormal{B}} & \mathbf{0}_{n_\textnormal{B}\times(n_\textnormal{T}-n_\textnormal{B})}\\
	\mathbf{0}_{(n_\textnormal{T}-n_\textnormal{B})\times n_\textnormal{B}} & \mathbf{0}_{(n_\textnormal{T}-n_\textnormal{B})\times(n_\textnormal{T}-n_\textnormal{B})}
	\end{array}\right)$,
$\tau=10^{-5}$, {and $\gamma^{(0)} = 1$. During each iteration, we update the sequence $\{\gamma^{(k)}\}$ by:}\footnote{It should be noted that $\mathbb{W}^{(0)}$ is not necessarily a feasible point in $\mathcal{W}_i$, $i\in\{1,2\}$. This is because the iteration variable $\mathbb{W}^{(1)}$ must be feasible by \eqref{eq: Wk update} since $\gamma^{(0)} = 1$.}
\begin{IEEEeqnarray}{rCl}
	\gamma^{(k+1)} =  \gamma^{(k)} \Bigl( 1-10^{-2} \gamma^{(k)} \Bigr). \label{eq: gamma sequence}
\end{IEEEeqnarray}
The algorithm is terminated if any of the following conditions is satisfied:
\begin{IEEEeqnarray}{rCl}
\left\{ \begin{array}{l}
	\|\mathbb{W}^{(k+1)} - \mathbb{W}^{(k)}\|_\mathsf{F} \leq 10^{-6},\\
	| f( \mathbb{W}^{(k+1)} ) - f( \mathbb{W}^{(k)} ) | \leq 10^{-6}.
\end{array}\right.
\end{IEEEeqnarray}
Figs.~\ref{fig: Nt4Nb1Ne1_full} and \ref{fig: Nt4Nb2Ne2_full} illustrate the secrecy rates achieved by the optimal fully-connected beamformer $\mathbb{W}_\textnormal{FC}^\star$ in $\mathcal{P}_1$ and the optimal fully-connected ZF beamformer $\mathbb{W}_\textnormal{ZF}^\star$ in $\mathcal{P}_2$. Here, an MISO channel corresponds to Fig.~\ref{fig: Nt4Nb1Ne1_full}, and an MIMO channel corresponds to Fig.~\ref{fig: Nt4Nb2Ne2_full}. To assess the effectiveness of the proposed beamforming design, the achievable secrecy rates are compared with those derived in Thm.~\ref{thm1}, which corresponds to a direct-connected scheme without beamforming.

\begin{figure}[t]
	\centering
	\includegraphics[width=3.5in]{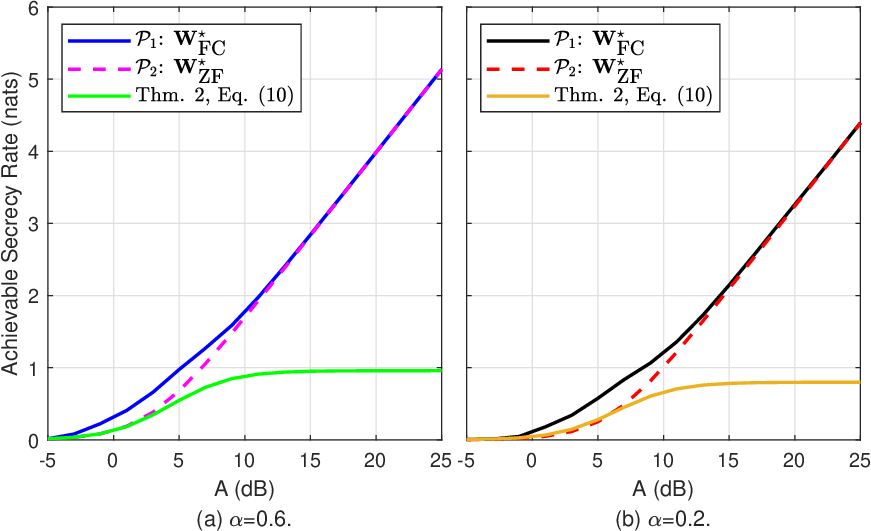}
	\caption{Achievable secrecy rates of fully-connected beamformer when $\mathbb{H}_\textnormal{B}=\mathbb{H}_\textnormal{B}^{1\times4}$, $\mathbb{H}_\textnormal{E}=\mathbb{H}_\textnormal{E}^{1\times4}$, and $\bm{\alpha}=\alpha\mathbf{1}_{4}$.}
	\label{fig: Nt4Nb1Ne1_full}
\end{figure}
\begin{figure}[t]
	\centering
	\includegraphics[width=3.5in]{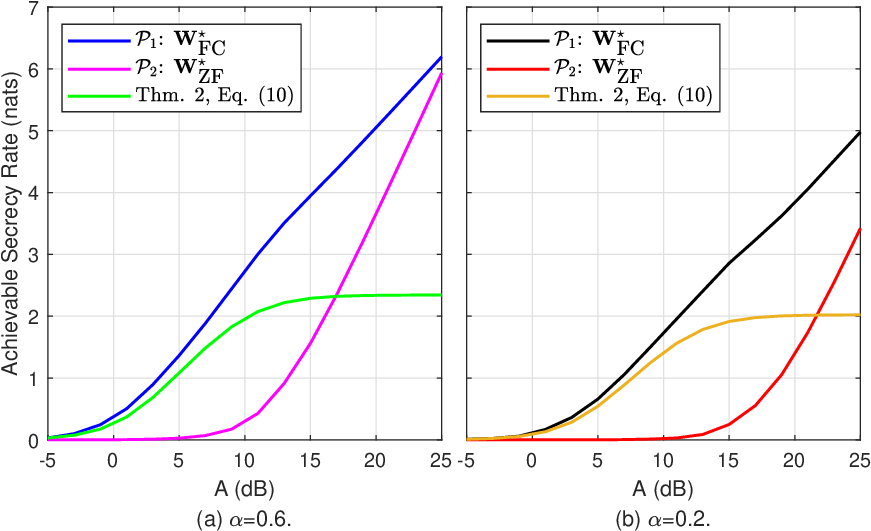}
	\caption{Achievable secrecy rates of fully-connected beamformer when $\mathbb{H}_\textnormal{B}=\mathbb{H}_\textnormal{B}^{2\times4}$, $\mathbb{H}_\textnormal{E}=\mathbb{H}_\textnormal{E}^{2\times4}$, and $\bm{\alpha}=\alpha\mathbf{1}_{4}$.}
	\label{fig: Nt4Nb2Ne2_full}
\end{figure}

As shown in Fig.~\ref{fig: Nt4Nb1Ne1_full}, both $\mathbb{W}_\textnormal{FC}^\star$ and $\mathbb{W}_\textnormal{ZF}^\star$ significantly enhance the secrecy rate compared to the direct-connected beamformer, thereby highlighting the potential secrecy gains through beamforming in practical VLC systems. Moreover, Fig.~\ref{fig: Nt4Nb1Ne1_full} reveals that for the MISO-VLC wiretap channel, $\mathbb{W}_\textnormal{FC}^\star$ and $\mathbb{W}_\textnormal{ZF}^\star$ exhibit nearly identical performance at high SNR, which is consistent with the observations in \cite{Mostafa2015}. From Fig.~\ref{fig: Nt4Nb2Ne2_full}, it is further observed that imposing the ZF constraint leads to a slight secrecy performance degradation compared with the unconstrained fully-connected scheme. Nevertheless, the ZF beamformer still substantially outperforms the direct-connected beamformer at high SNR, thereby demonstrating the effectiveness of the proposed ZF beamforming scheme. 


\subsubsection{Sub-connected Beamforming Schemes in $\mathcal{P}_3$, $\mathcal{P}_4$ and $\mathcal{P}_5$}\label{sec: sub-connect simulation}
We apply Algorithm~\ref{alg: alg2} to solve the sub-connected beamforming optimization problems $\mathcal{P}_3$ and $\mathcal{P}_4$, whereas the convex problem $\mathcal{P}_5$ is solved using CVX. The algorithm is initialized by: $\mathbb{B}^{(0)}=\mathbf{0}_{ n_\textnormal{B} \times ( n_\textnormal{T} -n_\textnormal{B} ) }$, and terminated if any of the following conditions is satisfied:
\begin{IEEEeqnarray}{rCl}
\left\{ \begin{array}{l}
	\|\mathbb{B}^{(k+1)} - \mathbb{B}^{(k)}\|_\mathsf{F} \leq 10^{-6},\\
	| f( \mathbb{B}^{(k+1)} ) - f( \mathbb{B}^{(k)} ) | \leq 10^{-6}.
\end{array}\right.
\end{IEEEeqnarray}
Figs.~\ref{fig: Nt4Nb1Ne1_sub} and \ref{fig: Nt4Nb2Ne2_sub} depict the secrecy rates achieved by the optimal sub-connected beamformer $\mathbb{B}_\textnormal{SC}^\star$ in $\mathcal{P}_3$, the optimal sub-connected ZF beamformer $\mathbb{B}_\textnormal{ZF}^\star$ in $\mathcal{P}_4$, and the optimal sub-connected MLSE beamformer $\mathbb{B}_\textnormal{MLSE}^\star$ in $\mathcal{P}_5$. Here, Fig.~\ref{fig: Nt4Nb1Ne1_sub} refers to an MISO channel, and Fig.~\ref{fig: Nt4Nb2Ne2_sub} refers to an MIMO channel. 

\begin{figure}[t]
	\centering
	\includegraphics[width=3.5in]{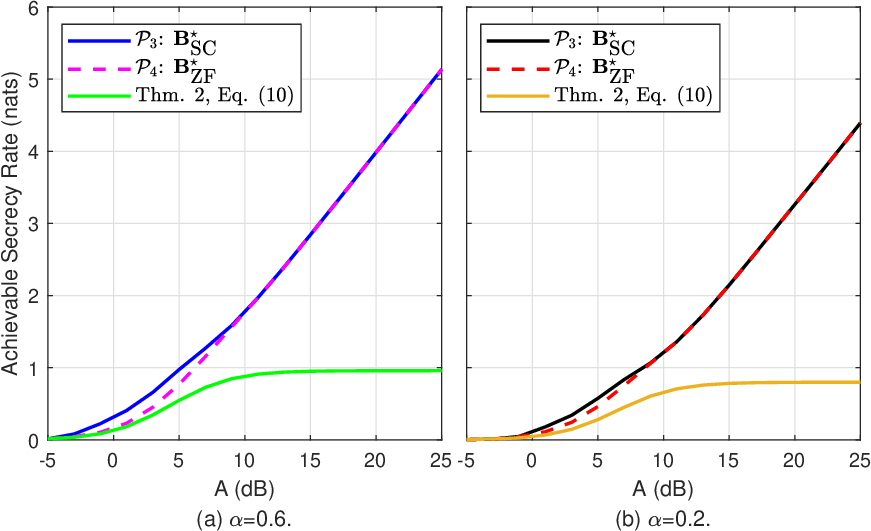}
	\caption{Achievable secrecy rates of sub-connected beamformer when $\mathbb{H}_\textnormal{B}=\mathbb{H}_\textnormal{B}^{1\times4}$, $\mathbb{H}_\textnormal{E}=\mathbb{H}_\textnormal{E}^{1\times4}$, and $\bm{\alpha}=\alpha\mathbf{1}_{4}$.}
	\label{fig: Nt4Nb1Ne1_sub}
	\vspace{-2mm}
\end{figure}

\begin{figure}[t]
	\centering
	\includegraphics[width=3.5in]{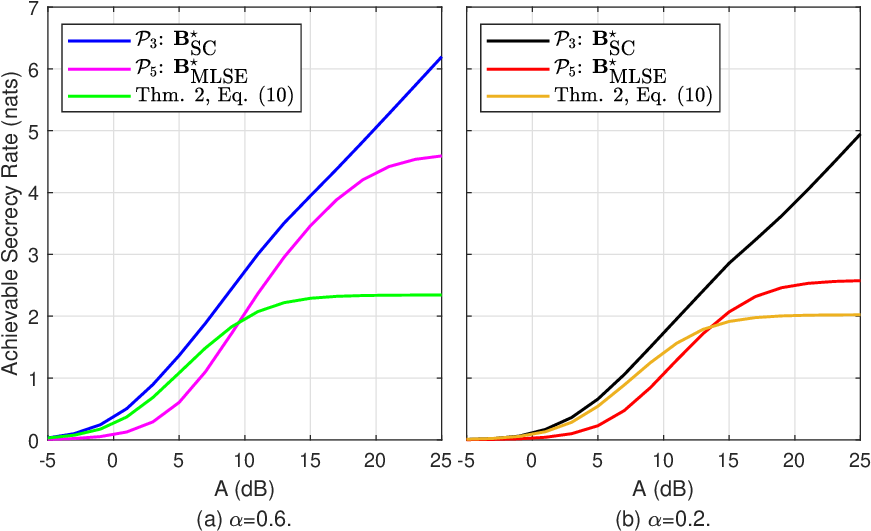}
	\caption{Achievable secrecy rates of sub-connected beamformer when $\mathbb{H}_\textnormal{B}=\mathbb{H}_\textnormal{B}^{2\times4}$, $\mathbb{H}_\textnormal{E}=\mathbb{H}_\textnormal{E}^{2\times4}$, and $\bm{\alpha}=\alpha\mathbf{1}_{4}$.}
	\label{fig: Nt4Nb2Ne2_sub}
\end{figure}

\begin{figure}[t]
	\centering
	\includegraphics[width=3.5in]{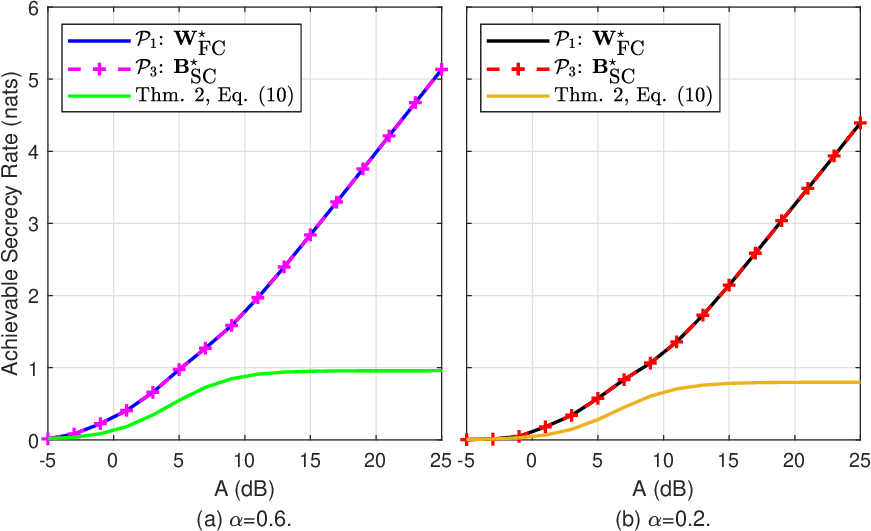}
	\caption{Comparison between fully- and sub-connected beamformers when $\mathbb{H}_\textnormal{B}=\mathbb{H}_\textnormal{B}^{1\times4}$, $\mathbb{H}_\textnormal{E}=\mathbb{H}_\textnormal{E}^{1\times4}$, and $\bm{\alpha}=\alpha\mathbf{1}_{4}$.}
	\label{fig: Nt4Nb1Ne1_fullandsub}
	\vspace{-2mm}
\end{figure}

\begin{figure}[t]
	\centering
	\includegraphics[width=3.5in]{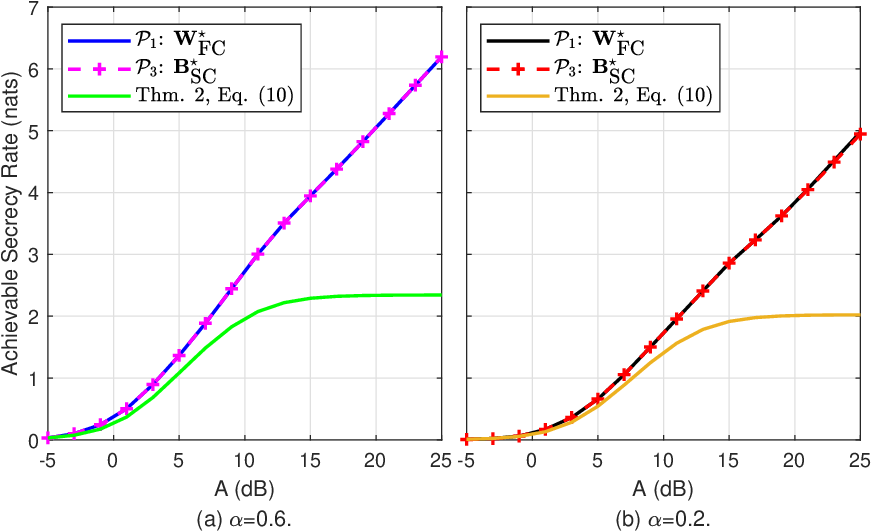}
	\caption{Comparison between fully- and sub-connected beamformers when $\mathbb{H}_\textnormal{B}=\mathbb{H}_\textnormal{B}^{2\times4}$, $\mathbb{H}_\textnormal{E}=\mathbb{H}_\textnormal{E}^{2\times4}$, and $\bm{\alpha}=\alpha\mathbf{1}_{4}$.}
	\label{fig: Nt4Nb2Ne2_fullandsub}
\end{figure}

As shown in Fig.~\ref{fig: Nt4Nb1Ne1_sub}, both $\mathbb{B}_\textnormal{SC}^\star$ and $\mathbb{B}_\textnormal{ZF}^\star$ achieve notable improvements in secrecy rate over the direct-connected beamformer, thereby confirming the effectiveness of sub-connected beamforming for enhancing physical-layer security. Similar to observations in Fig.~\ref{fig: Nt4Nb1Ne1_full}, $\mathbb{B}_\textnormal{SC}^\star$ and $\mathbb{B}_\textnormal{ZF}^\star$ achieve identical secrecy performance at high SNR for the MISO channel, indicating that the ZF constraint incurs no performance loss asymptotically. From Fig.~\ref{fig: Nt4Nb2Ne2_sub}, we observe that for the MIMO-VLC wiretap channel, the MLSE-based beamformer $\mathbb{B}_\textnormal{MLSE}^\star$ consistently outperforms the direct-connected beamformer as the SNR increases. This result highlights the advantage of incorporating MLSE beamforming in sub-connected architectures, particularly in scenarios where the ZF structure is impractical due to constraints.

\subsubsection{Comparison between Fully-connected and Sub-connected Beamforming Schemes in $\mathcal{P}_1$ and $\mathcal{P}_3$}\label{sec: comparison simulation}
Figs.~\ref{fig: Nt4Nb1Ne1_fullandsub} and \ref{fig: Nt4Nb2Ne2_fullandsub} compare the secrecy rates achieved by the optimal fully-connected beamformer $\mathbb{W}_\textnormal{FC}^\star$ and the optimal sub-connected beamformer $\mathbb{B}_\textnormal{SC}^\star$. The results show that $\mathbb{B}_\textnormal{SC}^\star$ achieves secrecy performance comparable to $\mathbb{W}_\textnormal{FC}^\star$ across a wide SNR range. This observation suggests that the sub-connected scheme can serve as a viable alternative to the fully-connected scheme, offering a favorable trade-off between secrecy performance and implementation complexity. With significantly reduced hardware connectivity at the transmitter, the sub-connected scheme is particularly attractive for practical VLC systems. These results further highlight the potential of sub-connected topologies for the secure VLC system design.


\subsubsection{Fully-connected Beamforming Scheme in $\mathcal{P}_6$}
Simulation parameters follow those used in Sec. \ref{sec: full-connect simulation}, with the exception that the algorithm is terminated if any of the following conditions is satisfied:
\begin{IEEEeqnarray}{rCl}
\left\{ \begin{array}{l}
	\|\mathbb{W}^{(k+1)} - \mathbb{W}^{(k)}\|_\mathsf{F} \leq 10^{-6},\\
	| f( \mathbb{W}^{(k+1)} ) - f( \mathbb{W}^{(k)} ) | \leq 10^{-6}.
\end{array}\right.
\end{IEEEeqnarray}
Figs.~\ref{fig: Nt1Nb4Ne4_full} and \ref{fig: Nt2Nb4Ne4_full} depict the secrecy rates achieved by the optimal fully-connected beamformer $\mathbb{W}_\textnormal{FC}^\star$, where an SIMO channel is considered in Fig.~\ref{fig: Nt1Nb4Ne4_full}, and an MIMO channel is considered in Fig.~\ref{fig: Nt2Nb4Ne4_full}. For comparison, the secrecy rates derived in Thm.~\ref{thm1} for the direct-connected scheme are also included. From these figures, it can be observed that the secrecy performance with beamforming is nearly identical to that achieved without beamforming. This result indicates that under Case II, the deployment of beamformer brings negligible secrecy performance gains. Consequently, beamforming can be safely omitted in this case, leading to a substantial reduction in implementation complexity without sacrificing secrecy performance in practical VLC systems.
\begin{figure}[t]
	\centering
	\includegraphics[width=3.5in]{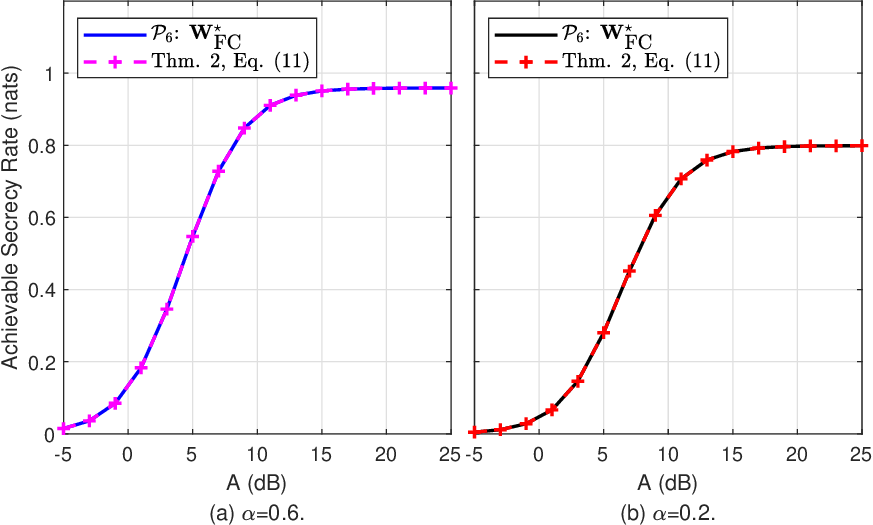}
	\caption{Achievable secrecy rates of fully-connected beamformer when $\mathbb{H}_\textnormal{B}=(\mathbb{H}_\textnormal{B}^{1\times4})^\mathsf{T}$, $\mathbb{H}_\textnormal{E}=(\mathbb{H}_\textnormal{E}^{1\times4})^\mathsf{T}$, and $\bm{\alpha}=\alpha$.}
	\label{fig: Nt1Nb4Ne4_full}
\end{figure}
 
\begin{figure}[t]
	\centering
	\includegraphics[width=3.5in]{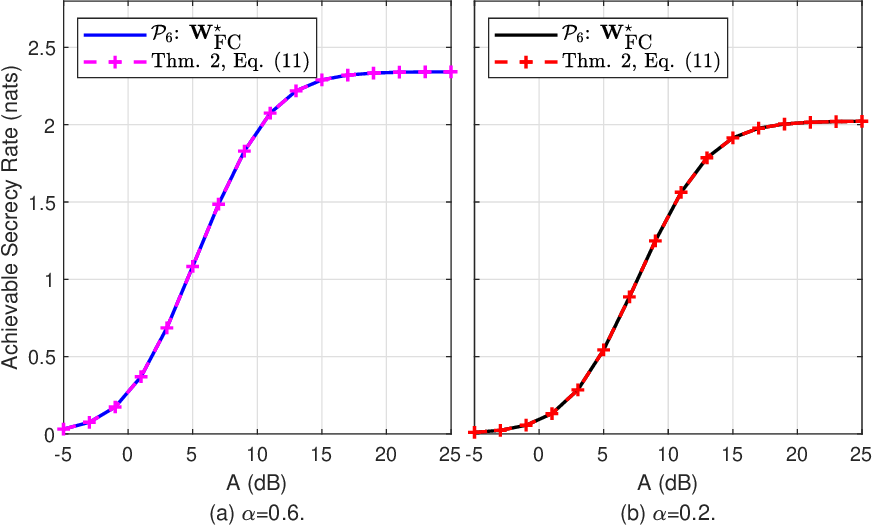}
	\caption{Achievable secrecy rates of fully-connected beamformer when $\mathbb{H}_\textnormal{B}=(\mathbb{H}_\textnormal{B}^{2\times4})^\mathsf{T}$, $\mathbb{H}_\textnormal{E}=(\mathbb{H}_\textnormal{E}^{2\times4})^\mathsf{T}$, and $\bm{\alpha}=\alpha\mathbf{1}_{2}$.}
	\label{fig: Nt2Nb4Ne4_full}
\end{figure}
\section{Conclusion}\label{sec: conclusion}
This paper investigates the secrecy capacity of the MIMO-VLC wiretap channel under both peak- and average-intensity constraints. Two representative quantitative relationships between the transmit and receive apertures are considered: one in which the number of LEDs is larger than or equal to that of the PDs, and another in which it is not. Assuming a truncated exponential input distribution, lower bounds on the secrecy capacity are derived by leveraging the EPI, the GEPI, and the maximum entropy principle. Building upon these theoretical results, beamforming designs are further studied to maximize the achievable secrecy rate for the MIMO-VLC wiretap channels. Although the resulting optimization problem is non-convex, an effective SCA-aided algorithm is developed to obtain efficient solution. Simulation results demonstrate the effectiveness of the proposed beamforming schemes. In particular, when the number of LEDs is larger than or equal to that of the PDs, the proposed beamformers can significantly improve the secrecy performance of the MIMO-VLC wiretap channel, whereas such performance gains diminish when the number of LEDs is less than that of the PDs.

\appendices
\section{Proof of Lemma~\ref{lemma 2}} \label{sec: appendix A}
We first derive the gradient and then the Hessian matrix of $f_\textnormal{B}(\mathbb{W})$ in the following \cite{Zhang2017}. Note that the differential of $f_\textnormal{B}(\mathbb{W})$ is
\begin{IEEEeqnarray}{rCl}
\mathrm{d}f_\textnormal{B}
&=& \mathrm{d} \left\{ -\frac{n_\textnormal{B}}{2} \log\left( 1 + |\mathbb{H}_\textnormal{B} \mathbb{W}^\mathsf{T} \cdot \diag{\mathbf{p}} \cdot \mathbb{W} \mathbb{H}_\textnormal{B}^\mathsf{T} |^{\frac{1}{n_\textnormal{B}}} \right) \right\} \nonumber\\
&=& -\frac{n_\textnormal{B}}{2} \frac{1}{ 1 + |\mathbb{H}_\textnormal{B} \mathbb{W}^\mathsf{T} \cdot \diag{\mathbf{p}} \cdot \mathbb{W} \mathbb{H}_\textnormal{B}^\mathsf{T} |^{\frac{1}{n_\textnormal{B}}} }\nonumber\\
&& \cdot \frac{1}{n_\textnormal{B}} |\mathbb{H}_\textnormal{B} \mathbb{W}^\mathsf{T} \cdot \diag{\mathbf{p}} \cdot \mathbb{W} \mathbb{H}_\textnormal{B}^\mathsf{T} |^{ \frac{1}{n_\textnormal{B}}-1 } \nonumber\\
&& \cdot \mathrm{d} |\mathbb{H}_\textnormal{B} \mathbb{W}^\mathsf{T} \cdot \diag{\mathbf{p}} \cdot \mathbb{W} \mathbb{H}_\textnormal{B}^\mathsf{T} | \\
&=& -\frac{1}{2} \frac{ |\mathbb{H}_\textnormal{B} \mathbb{W}^\mathsf{T} \cdot \diag{\mathbf{p}} \cdot \mathbb{W} \mathbb{H}_\textnormal{B}^\mathsf{T} |^{\frac{1}{n_\textnormal{B} }} }{ 1 + |\mathbb{H}_\textnormal{B} \mathbb{W}^\mathsf{T} \cdot \diag{\mathbf{p}} \cdot \mathbb{W} \mathbb{H}_\textnormal{B}^\mathsf{T} |^{\frac{1}{n_\textnormal{B}}} } \mathrm{tr}\Bigl\{ \bigl( \mathbb{H}_\textnormal{B} \mathbb{W}^\mathsf{T} \nonumber\\
&& \cdot \diag{\mathbf{p}} \cdot \mathbb{W} \mathbb{H}_\textnormal{B}^\mathsf{T} \bigr)^{-1} \mathrm{d} \bigl( \mathbb{H}_\textnormal{B} \mathbb{W}^\mathsf{T} \cdot \diag{\mathbf{p}} \cdot \mathbb{W} \mathbb{H}_\textnormal{B}^\mathsf{T} \bigr) \Bigr\} \label{eq: d|X|}\qquad\\
&=& -\frac{ |\mathbb{H}_\textnormal{B} \mathbb{W}^\mathsf{T} \cdot \diag{\mathbf{p}} \cdot \mathbb{W} \mathbb{H}_\textnormal{B}^\mathsf{T} |^{\frac{1}{n_\textnormal{B} }} }{ 1 + |\mathbb{H}_\textnormal{B} \mathbb{W}^\mathsf{T} \cdot \diag{\mathbf{p}} \cdot \mathbb{W} \mathbb{H}_\textnormal{B}^\mathsf{T} |^{\frac{1}{n_\textnormal{B}}} } \mathrm{tr}\Bigl\{ \mathbb{H}_\textnormal{B}^\mathsf{T} \bigl( \mathbb{H}_\textnormal{B} \mathbb{W}^\mathsf{T} \nonumber\\
&& \cdot \diag{\mathbf{p}} \cdot \mathbb{W} \mathbb{H}_\textnormal{B}^\mathsf{T} \bigr)^{-1} \mathbb{H}_\textnormal{B} \mathbb{W}^\mathsf{T} \cdot \diag{\mathbf{p}} \cdot \mathrm{d}\mathbb{W} \Bigr\}, \label{eq: tr(AB)}
\end{IEEEeqnarray}
where \eqref{eq: d|X|} holds since $\mathrm{d} |\mathbb{X}| = |\mathbb{X}| \mathrm{tr}(\mathbb{X}^{-1} \mathrm{d} \mathbb{X} )$, and \eqref{eq: tr(AB)} holds since $\mathrm{tr}\{\mathbb{X}\mathbb{Y}\} = \mathrm{tr}\{\mathbb{Y}^\mathsf{T}\mathbb{X}^\mathsf{T}\} = \mathrm{tr}\{\mathbb{Y}\mathbb{X}\}$. By \eqref{eq: tr(AB)}, we can obtain the gradient of $f_\textnormal{B}(\mathbb{W})$ in \eqref{eq: f1 gradient}.

For the Hessian matrix of $f_\textnormal{B}(\mathbb{W})$, we have
\begin{small}
\begin{IEEEeqnarray}{rCl}
\IEEEeqnarraymulticol{3}{l}{%
	\mathrm{d}^2 f_\textnormal{B} 
}\nonumber\\ 
&=& \mathrm{d} \{ \mathrm{d} f_\textnormal{B} \} \nonumber \\
&\approx& {\scriptsize - \mathrm{d} \biggl\{ \mathrm{tr}\Bigl\{ \mathbb{H}_\textnormal{B}^\mathsf{T} \bigl( \mathbb{H}_\textnormal{B} \mathbb{W}^\mathsf{T} \cdot \diag{\mathbf{p}} \cdot \mathbb{W} \mathbb{H}_\textnormal{B}^\mathsf{T} \bigr)^{-1} \mathbb{H}_\textnormal{B} \mathbb{W}^\mathsf{T} \cdot \diag{\mathbf{p}} \cdot \mathrm{d}\mathbb{W} \Bigr\} \biggr\} }\nonumber\\
&=& {\scriptsize -\mathrm{tr} \biggl\{ \mathrm{d}\Bigl\{ \mathbb{H}_\textnormal{B}^\mathsf{T} \bigl( \mathbb{H}_\textnormal{B} \mathbb{W}^\mathsf{T} \cdot \diag{\mathbf{p}} \cdot \mathbb{W} \mathbb{H}_\textnormal{B}^\mathsf{T} \bigr)^{-1} \mathbb{H}_\textnormal{B} \mathbb{W}^\mathsf{T}  \cdot \diag{\mathbf{p}} \cdot \mathrm{d}\mathbb{W} \Bigr\} \biggr\} } \nonumber\\
&=& -\mathrm{tr} \Bigl\{ \mathbb{H}_\textnormal{B}^\mathsf{T} \cdot \mathrm{d} \bigl( \mathbb{H}_\textnormal{B} \mathbb{W}^\mathsf{T} \cdot \diag{\mathbf{p}} \cdot \mathbb{W} \mathbb{H}_\textnormal{B}^\mathsf{T} \bigr)^{-1} \mathbb{H}_\textnormal{B} \mathbb{W}^\mathsf{T} \cdot \diag{\mathbf{p}} \cdot \mathrm{d}\mathbb{W} \Bigr\} \nonumber\\
&& -\mathrm{tr} \Bigl\{ \mathbb{H}_\textnormal{B}^\mathsf{T} \bigl( \mathbb{H}_\textnormal{B} \mathbb{W}^\mathsf{T} \cdot \diag{\mathbf{p}} \cdot \mathbb{W} \mathbb{H}_\textnormal{B}^\mathsf{T} \bigr)^{-1} \mathbb{H}_\textnormal{B} (\mathrm{d} \mathbb{W}^\mathsf{T}) \cdot \diag{\mathbf{p}} \cdot \mathrm{d}\mathbb{W} \Bigr\}. \nonumber\\ \label{eq: d2 f1 1}
\end{IEEEeqnarray}
\end{small}
Note that 
\begin{small}
\begin{IEEEeqnarray}{rCl}
\IEEEeqnarraymulticol{3}{l}{%
	\mathrm{d} \bigl( \mathbb{H}_\textnormal{B} \mathbb{W}^\mathsf{T} \cdot \diag{\mathbf{p}} \cdot \mathbb{W} \mathbb{H}_\textnormal{B}^\mathsf{T} \bigr)^{-1}
}\nonumber\\
&=& -\bigl( \mathbb{H}_\textnormal{B} \mathbb{W}^\mathsf{T} \cdot \diag{\mathbf{p}} \cdot \mathbb{W} \mathbb{H}_\textnormal{B}^\mathsf{T} \bigr)^{-1} \mathrm{d} \bigl( \mathbb{H}_\textnormal{B} \mathbb{W}^\mathsf{T} \cdot \diag{\mathbf{p}} \cdot \mathbb{W} \mathbb{H}_\textnormal{B}^\mathsf{T} \bigr) \nonumber\\
&& \cdot \bigl( \mathbb{H}_\textnormal{B} \mathbb{W}^\mathsf{T} \cdot \diag{\mathbf{p}} \cdot \mathbb{W} \mathbb{H}_\textnormal{B}^\mathsf{T} \bigr)^{-1} \nonumber\\
&=& -\bigl( \mathbb{H}_\textnormal{B} \mathbb{W}^\mathsf{T} \cdot \diag{\mathbf{p}} \cdot \mathbb{W} \mathbb{H}_\textnormal{B}^\mathsf{T} \bigr)^{-1} \mathbb{H}_\textnormal{B} (\mathrm{d} \mathbb{W}^\mathsf{T} ) \cdot \diag{\mathbf{p}} \cdot \mathbb{W} \mathbb{H}_\textnormal{B}^\mathsf{T} \nonumber\\
&& \cdot \bigl( \mathbb{H}_\textnormal{B} \mathbb{W}^\mathsf{T} \cdot \diag{\mathbf{p}} \cdot \mathbb{W} \mathbb{H}_\textnormal{B}^\mathsf{T} \bigr)^{-1} -\bigl( \mathbb{H}_\textnormal{B} \mathbb{W}^\mathsf{T} \cdot \diag{\mathbf{p}} \cdot \mathbb{W} \mathbb{H}_\textnormal{B}^\mathsf{T} \bigr)^{-1} \nonumber\\
&& \cdot \mathbb{H}_\textnormal{B} \mathbb{W}^\mathsf{T} \cdot \diag{\mathbf{p}} \cdot (\mathrm{d} \mathbb{W}) \mathbb{H}_\textnormal{B}^\mathsf{T} \bigl( \mathbb{H}_\textnormal{B} \mathbb{W}^\mathsf{T} \cdot \diag{\mathbf{p}} \cdot \mathbb{W} \mathbb{H}_\textnormal{B}^\mathsf{T} \bigr)^{-1}. \nonumber\\ \label{eq: d2 f1 2}
\end{IEEEeqnarray}
\end{small}Substituting \eqref{eq: d2 f1 2} into \eqref{eq: d2 f1 1}, we can conclude the Hessian matrix of $f_\textnormal{B}(\mathbb{W})$ in \eqref{eq: f1 hessian}.


\end{document}